\documentclass[a4paper, 11pt]{article}

\usepackage[pdftex]{graphicx}
\usepackage[cmex10]{amsmath}
\usepackage{amssymb}
\usepackage{booktabs}
\usepackage{url}
\usepackage[margin=2cm]{geometry}

\newcommand{\bK}{\mathbf{K}}
\newcommand{\bY}{\mathbf{Y}}

\newcommand{\bX}{\mathbf{X}}
\newcommand{\bx}{\mathbf{x}}
\newcommand{\bbeta}{\boldsymbol \beta}

\newcommand{\beps}{\boldsymbol \varepsilon}
\newcommand{\real}{\mathbb{R}}
\newcommand{\mN}{\mathcal{N}}
\newcommand{\bT}{\mathbf{T}}

\begin{document}

\title{Probabilistic Forecasting of Temporal Trajectories of Regional Power Production---Part 2: Photovoltaic Solar}

\author{Thordis~L.~Thorarinsdottir,
Anders~L{\o}land, and
Alex~Lenkoski\thanks{Norwegian Computing Center, Oslo, Norway (e-mail: thordis@nr.no).}
}

\maketitle

\begin{abstract}
We propose a fully probabilistic prediction model for spatially aggregated solar photovoltaic (PV) power production at an hourly time scale with lead times up to several days using weather forecasts from numerical weather prediction systems as covariates. After an appropriate logarithmic transformation of the power production, we develop a multivariate Gaussian prediction model under a Bayesian inference framework. The model incorporates the temporal error correlation yielding physically consistent forecast trajectories. Several formulations of the correlation structure are proposed and investigated. Our method is one of a few approaches that issue full predictive distributions for PV power production. In a case study of PV power production in Germany, the method gives calibrated and skillful forecasts.  
\end{abstract}

\section{Introduction}

Part 1 of the paper proposed
 fully probabilistic prediction models for spatially aggregated
wind  power production at an hourly time
scale with lead times up to several days using weather forecasts from
numerical weather prediction systems as covariates.
This part of the paper (Part 2) is concerned with corresponding models for photovoltaic (PV) solar power production. The proposed PV models share the same framework as the wind power model from Part 1.

The increase in energy production from renewable energy sources is
driven by wind and PV power production.  In Germany, for instance, renewable
energy accounted for $36.0\%$ of the total national energy production in
2017 compared to $6.6\%$ in 2000 according to the Arbeitsgemeinschaft
Energiebilanzen, a working group founded by energy related
associations in Germany.  This increase is to a large extent due to
expansion in wind and photovoltaic (PV) solar power production. (PV
production accounted for $0.0\%$ in 2000 and increased to 
$6.6\%$ in 2017.)  As discussed in Part 1,
management of electricity grids, scheduling of the conventional
production and general energy  
market decisions call for a probabilistic forecasting framework
\cite{Gneiting2011}.  While probabilistic forecasts are becoming
increasingly frequent for wind power forecasting 
 \cite{Bremnes2004, PinsonKariniotakis2010, JeonTaylor2012,Pinson2012, Pinson2013,Hong&2016,Dowell&2016}
they have been less common for PV power forecasts \cite{Lorenz&2009,
  Bracale&2013,alessandrini2015analog,sperati2016application,golestaneh2016very} and only few approaches have issued full predictive
distributions.  However, the 
Global Energy Forecasting Competition
(GEFCom2014 \cite{Hong&2016}) has spurred probabilistic forecasts
based on non-parametric approaches
\cite{Mohammed&2015,nagy&2016,Huang&2016,agoua2018probabilistic,golestaneh2016very}.

While the wind power model we developed in Part 1 relied on wind speed
forecasts from numerical weather prediction (NWP) models, the
essential input for a NWP-based prediction model for PV solar power is a forecast of the downward solar radiation at the surface, which is also called global horizontal irradiance (GHI), see e.g.\ \cite{bacher2009} and \cite{MathiesenKleissl2011}.  An explicit physical model for a single unit may also include forecasts of cloud cover, temperature and snow cover \cite{lorenz2011, PellandGalanisKallos2013}.    

As for the wind model, we propose to directly predict the aggregate
country-wide PV production using spatially averaged NWP forecasts of
the relevant weather variables as inputs.  We specify the
probabilistic prediction model for PV power as a Bayesian hierarchical model, which allows us to incorporate a correlation structure in both the model parameters associated with each lead time  as well as the error structure across lead times.

The NWP forecasts and the German PV power production data are
introduced in Section~\ref{sec:data}.  The prediction models and the
statistical inference methods are presented in
Section~\ref{sec:models}.  The results, including forecast
verifications, are given in
Section~\ref{sec:results}. Section~\ref{sec:discussion} contains a final discussion.

\section{Forecast and observation data}\label{sec:data}

We employ the NWP forecast ensemble issued by the European Centre for Medium-Range Weather Forecasts (ECMWF), which has been shown to perform well for solar power \cite{MathiesenKleissl2011}.   The 50-member ECMWF ensemble system operates at a global horizontal resolution of $0.25 \times 0.25$ degrees, a resolution of approximately $32 \times 32$ km over Germany, and a temporal resolution of 3--6 h with lead times up to ten days \cite{LeutbecherPalmer2008, Molteni&1996}. We restrict attention to the forecast initialized at 00:00 UTC, corresponding to 2:00 am local time in summer and 1:00 am local time in winter, and lead times up to 72 h for accumulated global horizontal irradiance (GHI).  

The hourly solar power production data for Germany are obtained from the European Energy Exchange (EEX) where they are available to all members that trade on the EEX, see \url{www.transparency.eex.com/en/}.  We use data from the calendar year 2011 to assess the optimal length of the training period in the parameter estimation as well as for determining the prior parameters of the Bayesian model.  Given these values, we then test our methods on data from 2012.  In order to obtain equally long training periods for all dates, data from the previous year is used for the parameter estimation at the beginning of a year. 

We reduce the ECMWF ensemble to a single forecast given by the ensemble average.  For the operation and management of electricity grids, power production predictions are needed on an hourly basis.  However, for the first 72 h, the ECMWF forecasts have a temporal resolution of 3 h.  We therefore derive hourly forecasts through a spline interpolation conditional on the variables being non-negative.  A more advanced interpolation approach for GHI performs a temporal interpolation over the clear sky index, see e.g. \cite{Lorenz&2008}.  
In a third preprocessing step, we aggregate the forecasts in space by taking the spatial average.  For GHI, several studies have found that spatial averaging increases the skill of the forecast due to difficulty in dealing with changing cloud cover \cite{Girodo2006, Lorenz&2009, MathiesenKleissl2011}.  We employ here the average GHI forecast over all grid locations within Germany, resulting in a GHI forecast which is an average over 724 grid locations.

\section{PV solar power prediction model}\label{sec:models}

The downward solar radiation at the surface, also called global horizontal irradiance (GHI), is composed of the direct solar radiation at the surface and a sky diffusion component.  For an individual PV system, the two components of the GHI are used to generate a tilted forecast of irradiance in the plane of the PV arrays,  $x^*_{s}$ \cite{PellandGalanisKallos2013}.  Given $x^*_{s}$ and the local ambient temperature forecast, $x_a$, the power output of the PV system is then given by  $x^*_{s} f(x^*_{s}, x_a)$ for a non-linear function $f$ which parameters depend on the type of installment \cite{Drews&2007, lorenz2011, PellandGalanisKallos2013}.  Here, we model the aggregated power output as a function of the predicted GHI only, which is similar to the approach of \cite{bacher2009}.  Including temperature and snow depth as covariates did not improve the average marginal predictive performance (results not shown). Installing PV modules on northern latitudes can lead to snow losses of up to 20\%, but the effect of snow on individual PV modules can be predicted fairly well \cite{AndrewsPearce2012}. 

\begin{figure}[!hbpt]
\centering
\includegraphics[width=0.45\textwidth]{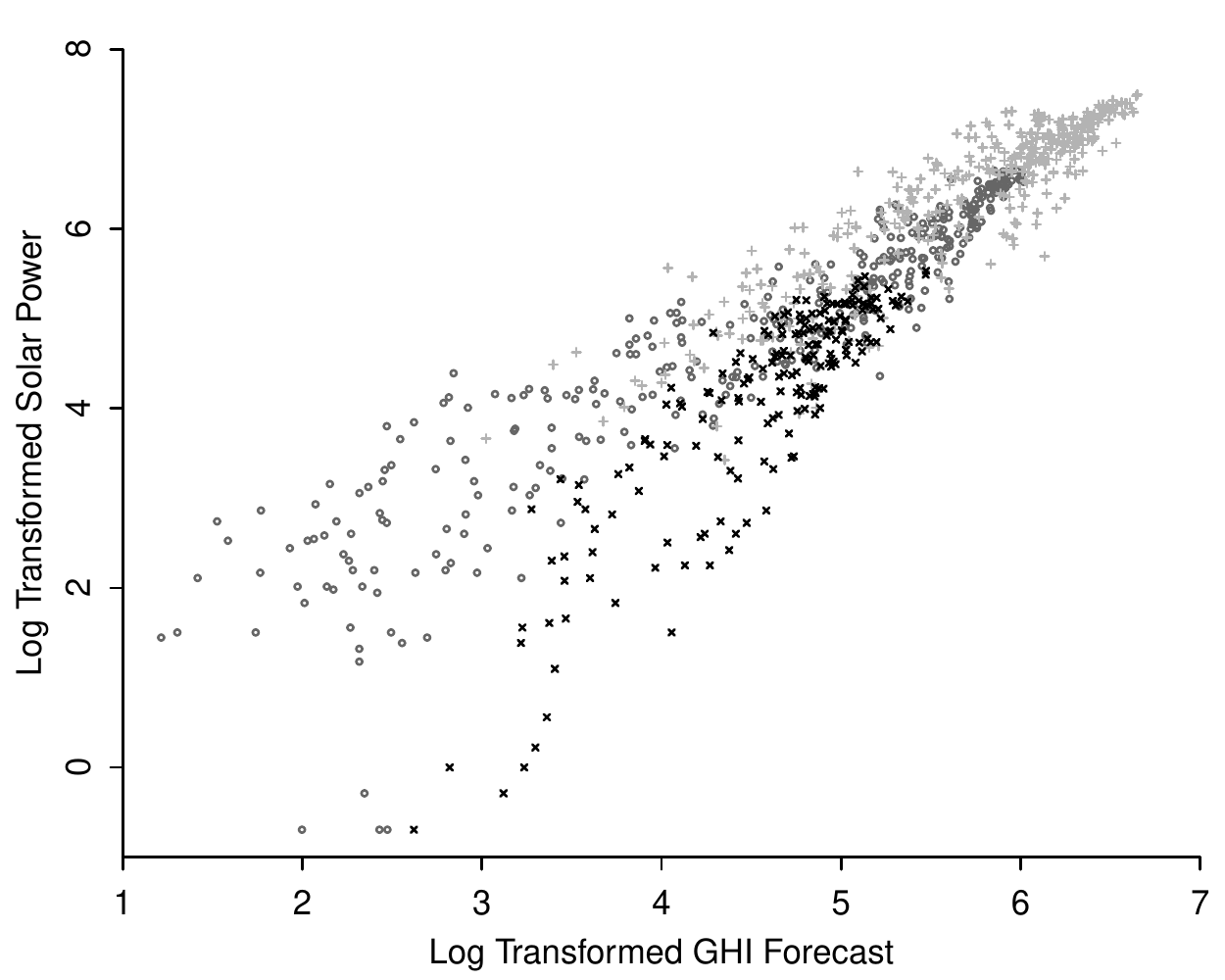}
\caption{Relationship between hourly solar power production in Germany in 2011 and the corresponding global horizontal irradiation (GHI) forecasts at 10:00 am (9:00 am) local time in summer (winter) ($\circ$), 4:00 pm (3:00 pm) local time in summer (winter) ($+$) and 9:00 pm (8:00 pm) local time in summer (winter) ($\times$). The plot only shows data points where both the power production and the GHI forecast are positive.}\label{fig:log sun}
\end{figure}

Denote by $x_{st}$ the GHI forecast for lead times $t$ and let $\tilde{y}_t$ denote the most recent available observed power production of the hour of the day that is being predicted at lead time $t$. That is, for $T = 72$ it holds that $\tilde{y}_t = \tilde{y}_{t + 24} = \tilde{y}_{t+48}$ for all $t \in \{1,\ldots, 24\}$.  We can now define
\[
\bT_+ := \{ t \in \{1, \ldots, T\} \, : \, \tilde{y}_t, x_{st} > 0\} 
\]
and 
\[
\bT_0 := \{ t \in \{1, \ldots, T\} \, : \, \tilde{y}_t = 0 \textup{ or } x_{st} = 0\}, 
\]
such that $\bT_+$ and $\bT_0$ are disjoint sets with $\bT_+ \cup \bT_0 = \{1,\ldots, T\}$. 
 The solar power production $Y_{st}$ at time $t \in \{t_1,\ldots,t_{| \bT_+|} \} = \bT_+$ is then given by 
\begin{equation}\label{eq:solar model}
\log( Y_{st}) = \beta_{0t} + \beta_{1t} \log(x_{st}) + \varepsilon_t,
\end{equation}
where $\beta_{0t}, \beta_{1t} \in \real$ and the error vector fulfills $\beps \sim \mN_{| \bT_+|} (0, \bK^{-1})$ for some precision matrix $\bK$.  As shown in Fig.~\ref{fig:log sun}, the relationship between the log transformed GHI forecast and the log transformed solar power production depends somewhat on the time of the day.  

We now set
\[
\bX = [\mathbb{I}_{|\bT_+|}~\textup{Diag}(\log(\bx_s))]
\]
with $\bx_s = (x_{st_1},\ldots,x_{st_{|\bT_+|}})^\top$ and, correspondingly, $\bY = (\log(Y_{s t_1}), \ldots, \log(Y_{s t_{|\bT_+|}}))^\top$.  The likelihood model for $\bY$ is then given by
\begin{equation}\label{eq:sun likelihood}
\bY \sim \mathcal{N}_{|\bT_+|}(\bX \bbeta, \, \bK^{-1}), 
\end{equation}
where $\bbeta = (\bbeta_0^\top, \bbeta_1^\top)^\top$ with $\bbeta_{i} = (\beta_{it_1}, \ldots, \beta_{it_{|\bT_+|}})^\top$ for $i = 0,1$. Inference is performed under a Bayesian paradigm in the same manner as described in Part 1. The regression parameters $\bbeta$ are given a normal conjugate prior distribution and the conjugate prior distribution for the precision matrix $\bK$ is the G-Wishart distribution $\bK \sim \mathcal{W}_G (3, \mathbb{I}_T)$ \cite{Roverato2002, Lenkoski2013}.  The support of $\mathcal{W}_G$ is the space of all symmetric positive definite matrices which fulfill the conditional independence structure given by the graph $G = (V,E)$ where $V = \{1, \ldots, |\bT_+|\}$ and $E \subset V \times V$.  That is, $K_{ij} = 0$ whenever $(i,j) \notin E$. For instance, if $G$ is the conditional independence structure of an autoregressive process of order $1$, AR(1), then it holds that $(i,j) \in E$ if and only if $|i -j| \leq 1$ and $|t_i -t_j| \leq 1$, such that correlation is only possible between consecutive time points. 

If the training data contains instances in which either the solar power production or the GHI forecast is equal to zero, we treat these as missing data.  For lead times $t \in \bT_0$, we follow \cite{PellandGalanisKallos2013} and set the predicted power production equal to the average observed production for this lead time during the training period. Typically, this value will be equal to zero. For instance, approximately 43\% of the observed hourly solar power production values in 2011 are equal to zero; for five night time hours no production is recorded throughout the entire year. 

\section{Results\label{sec:results}}

Here, we present assessments of marginal and multivariate predictive performance. The predictive performance is measured in terms of calibration and accuracy. A forecasting model is said to be calibrated if predicted probabilities are observed with the same relative frequency in the observations. This is assessed empirically through probability integral transform (PIT) histograms marginally and through band depth rank histograms in higher dimensions \cite{Dawid1984, Thorarinsdottir&2016}. For both cases, a uniform histogram indicates a calibrated forecast, while deviations from uniformity may provide information regarding the misspecification of the prediction model. Prediction accuracy is assessed by using proper scoring rules where a smaller score indicates a better performance with the errors given in the unit of the predictand \cite{GneitingRaftery2007}. Specifically, we apply the continuous ranked probability score (CRPS) which assesses the full predictive distribution as well as the mean absolute error (MAE) and the root mean squared error (RMSE) which assess the median and the mean of the predictive distribution, respectively. Further details on the forecast verification methods are given in Part 1.  

\subsection{Length of training period} 

\begin{figure}[!hbpt]
\centering
\includegraphics[width=0.5\textwidth]{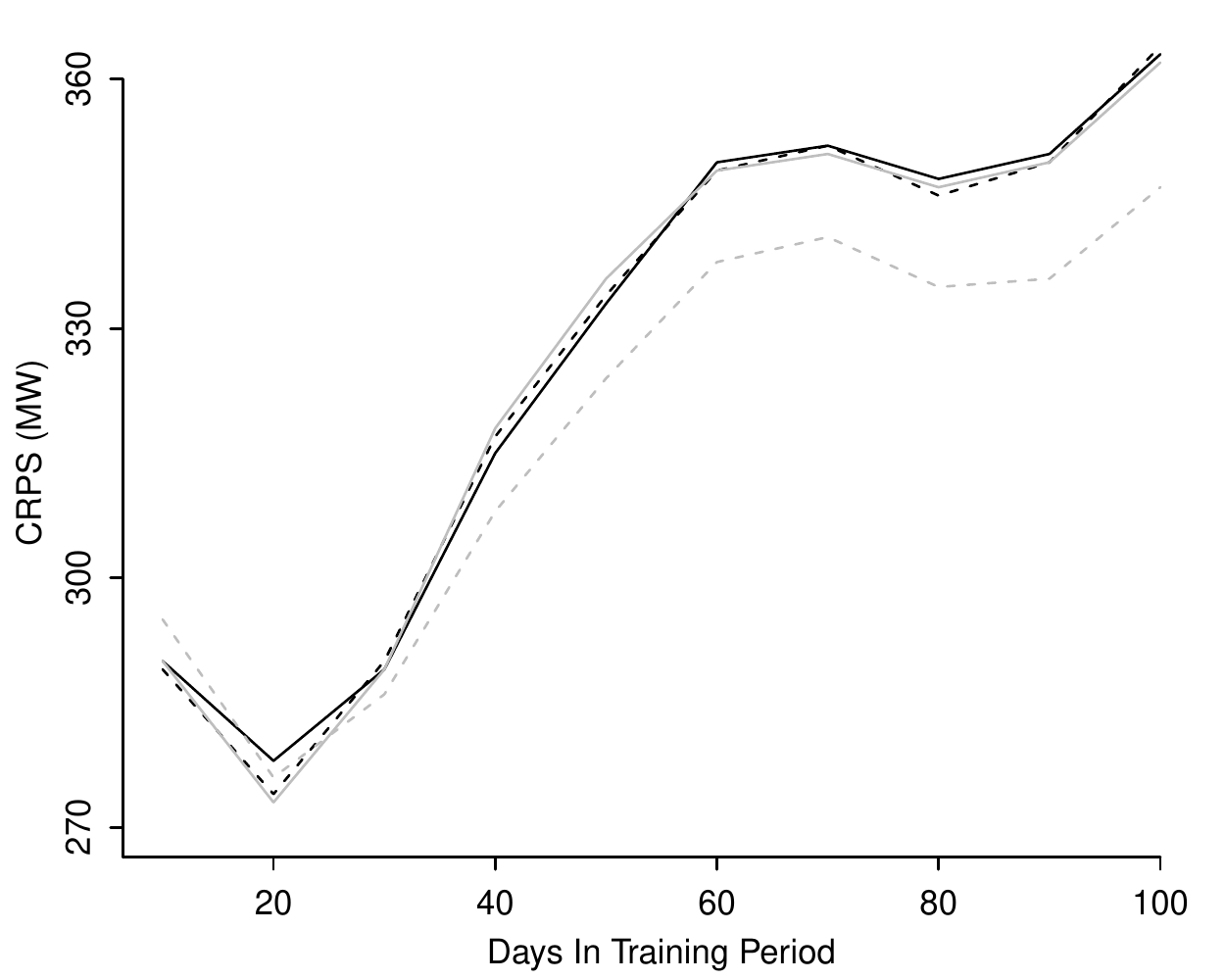}
\caption{Average continuous ranked probability score (CRPS) for solar power predictions aggregated over lead times up to 24 hours and the months of January, April, July and October of 2011 as a function of the number of days in the rolling training period. The plot shows the results for the full model (gray dashed line) as well as for independent marginal models with independence (black solid line), AR(1) (black dashed line), or AR(2) (gray solid line) structure on the regression coefficients. 
}\label{fig:sun training period}
\end{figure}

We assess the influence of the amount of training data on the results by comparing the average marginal predictive performance under rolling training periods of different lengths. 
The performance of the marginal prediction models for solar power
production is significantly more sensitive to the length of the
rolling training period than is the case for wind power, see
Fig.~\ref{fig:sun training period}. Results based on other
performance measures show a similar pattern.  Following these results,
we employ a rolling training period of length 20 days for the full
model and all marginal models.  For the multivariate copula models, we
observe nearly identical results when predicting daily sums and maxima
using rolling training periods of length 50 to 150 days. Shorter
training periods give somewhat worse performance for the maxima. We
thus use a training period of 100 days for estimating the multivariate
correlation structure in the copula models.  

\subsection{Marginal predictive performance}

\begin{figure}[!hbpt]
\centering
\includegraphics[width=0.25\textwidth]{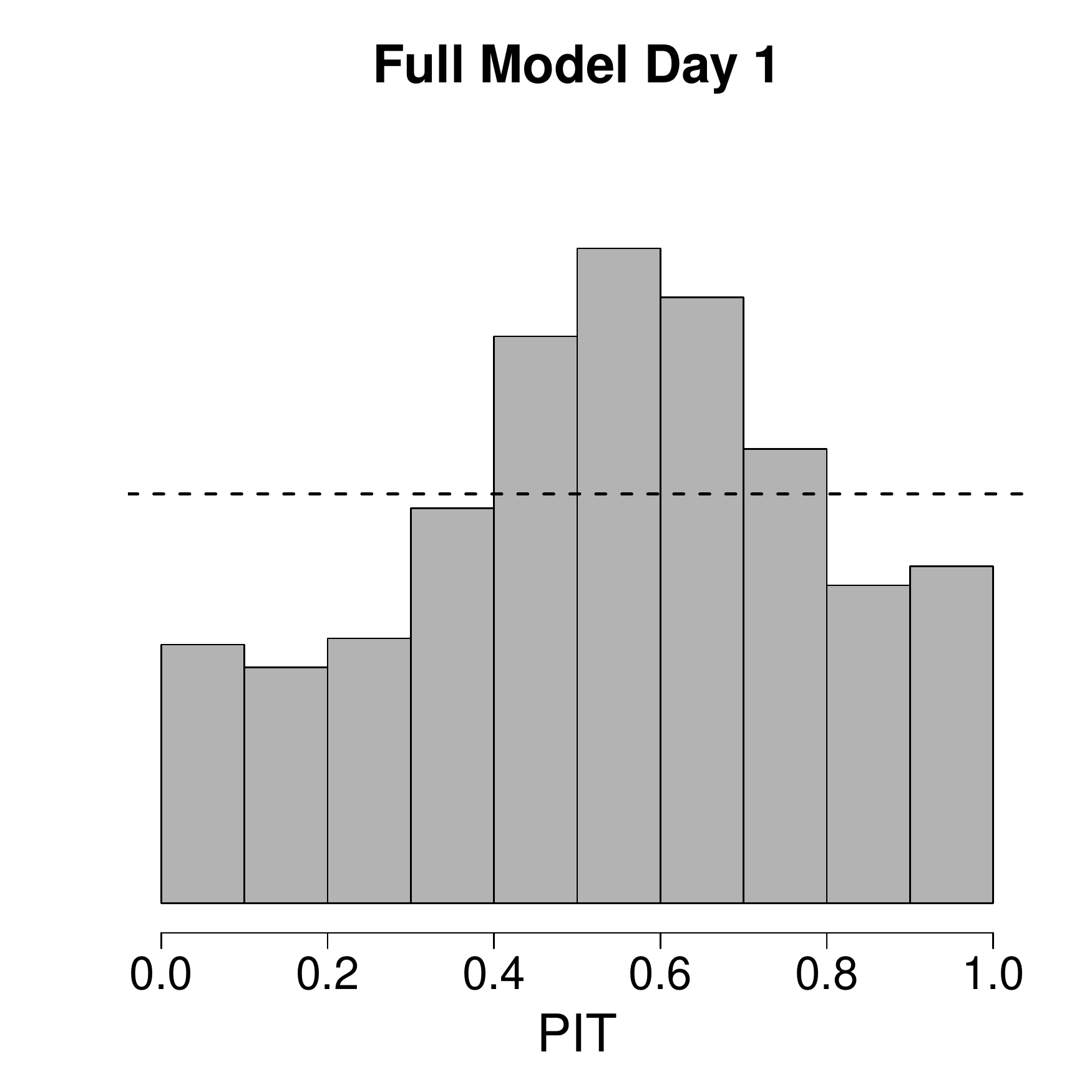}
\includegraphics[width=0.25\textwidth]{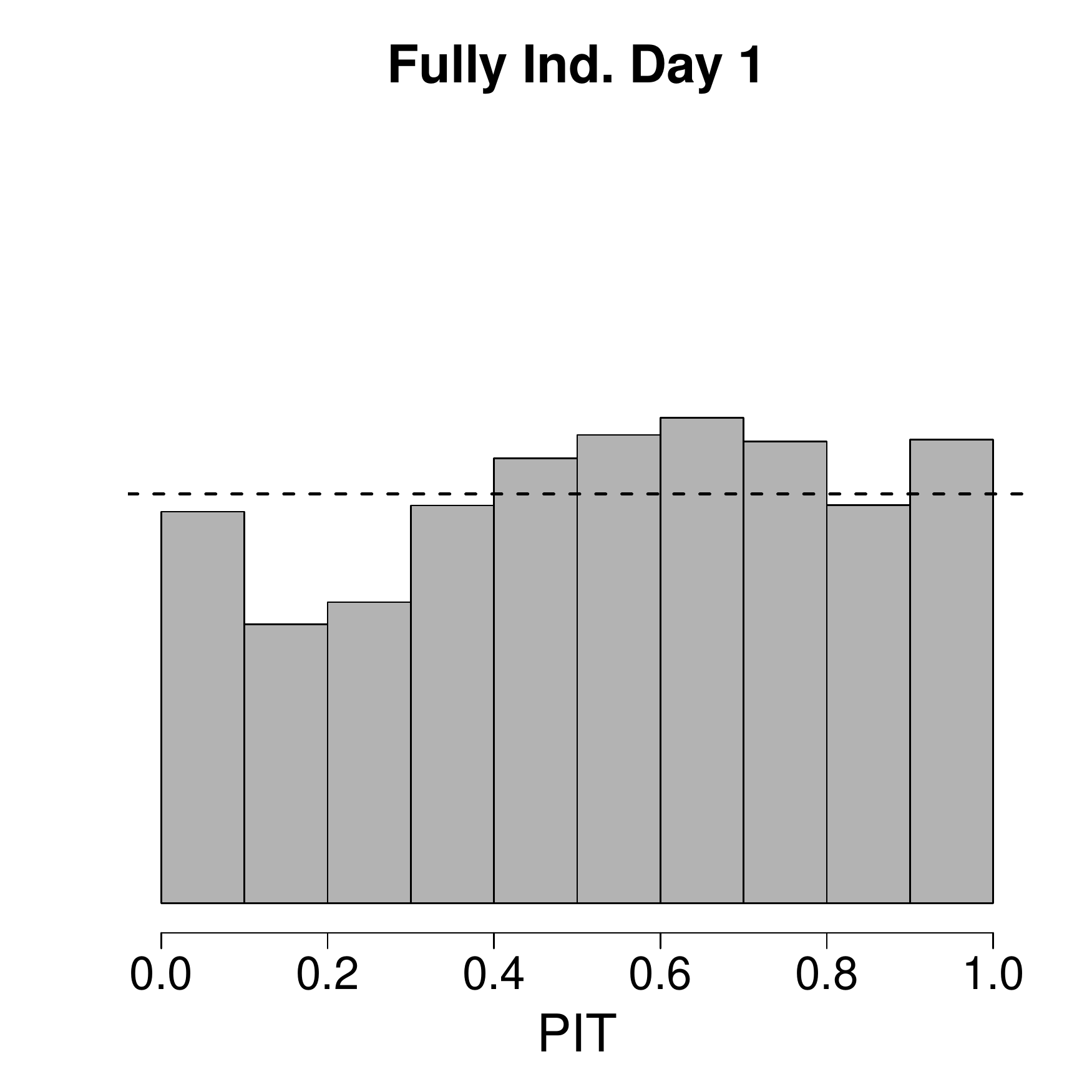}
\includegraphics[width=0.25\textwidth]{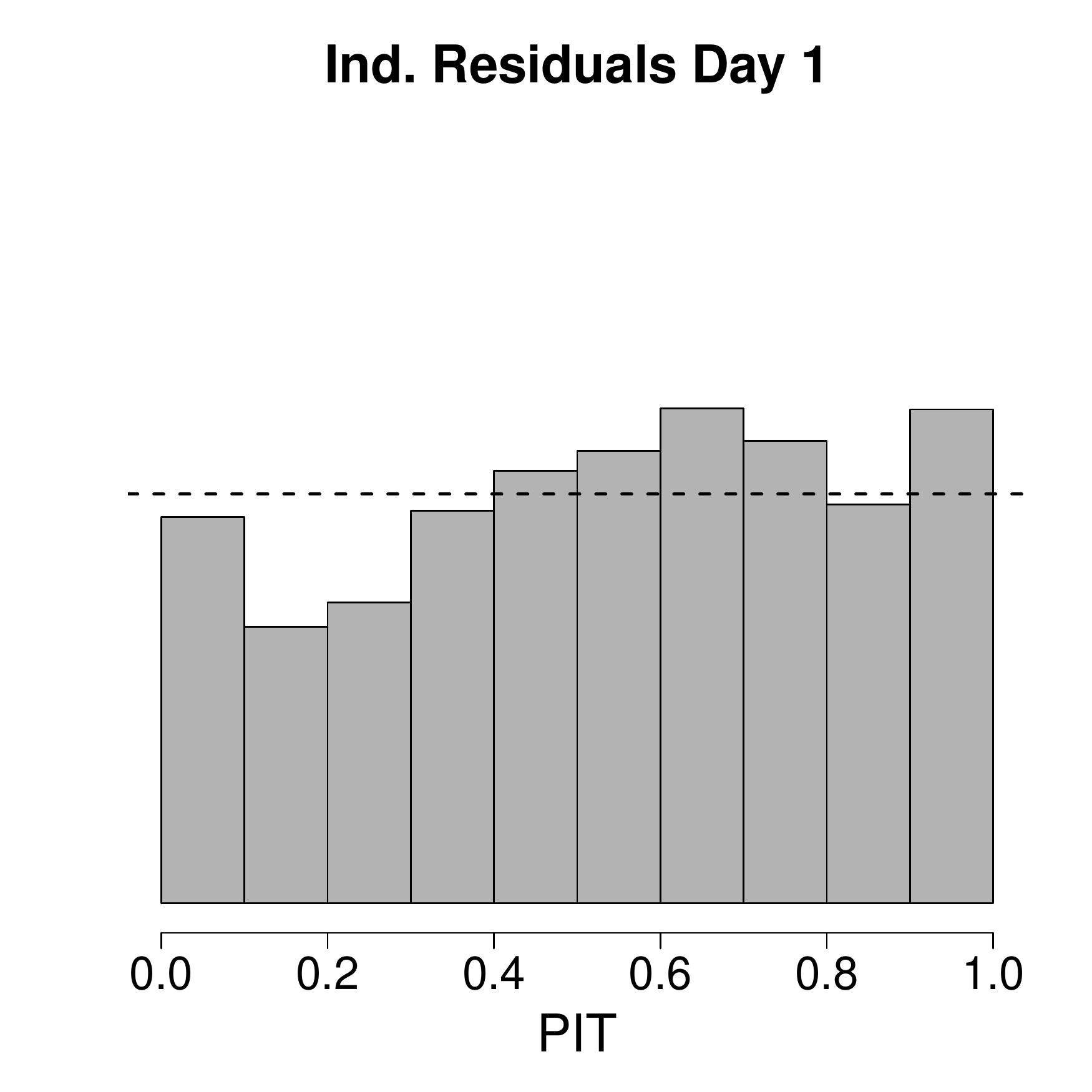}
\caption{Probility integral transform (PIT) histograms for marginal sun power predictions under three different marginal models. The PIT values are aggregated over lead times of 1-24h during the test period from January 1, 2011 to December 31, 2012 for which a probabilistic forecast was issued, a total of 9741 forecast cases. The dashed lines indicate the level of a perfectly flat histogram.}\label{fig:PIT marginal sun}
\end{figure}

We compare three models, (i) ``Full Model'' which has a AR(1) structure on the precision matrix for both the regression coefficients and residuals, (ii) ``Fully Independent'' which has fully independent residuals and regressions coefficients and ``Independent Residuals'' which models dependent regression coefficients but indepdent errors.  We start by assessing the marginal predictive performance of these three models.
The PIT histograms 
(Fig. \ref{fig:PIT marginal sun}) all indicate that the full model is over-dispersive, as the majority of observations fall in the middle quantiles of the distribution.  However, both the Fully Independent and Independent Residual models show some signs of slight upward bias but otherwise good calibration.

We further  measure calibration and sharpness of marginal predictions for sun
power by the width and coverage of 80\% prediction intervals. The
results are aggregated over lead times of 1-24h (Day 1), 25-48h (Day
2) and 49-72h (Day 3) in Table \ref{tab:calib}. The width of the
prediction intervals is roughly constant with respect to lead time, possibly due to the
systematic componenent in the GHI forecasts. The Independent Residuals model gives the narrowest prediction intervals,
while the Full Model has the best coverage, close to the desired 80\%.  However, this improved coverage comes at the expense of prediction intervals that are roughly 37\% larger than the Fully Independent or Independent Residual models.

\begin{table*}[ht]
\centering
\caption{Calibration and sharpness of marginal predictions for sun power as measured by the width and coverage of 80\% prediction intervals.  The results are aggregated over lead times of 1-24h (Day 1), 25-48h (Day 2) and 49-72h (Day 3), and the test period from January 1, 2011 to December 31, 2012. The best results in each category are indicated in italics.\label{tab:calib}}
\vspace{1mm}
\begin{tabular}{lcccccc}
\toprule
& \multicolumn{3}{c}{Width (MW)} & \multicolumn{3}{c}{Coverage (\%)} \\
\cmidrule(lr){2-4} \cmidrule(lr){5-7} 
& Day 1 & Day 2 & Day 3 & Day 1 & Day 2 & Day 3 \\ 
\midrule
Full Model & 5100 & 5099 & 5099 & \emph{0.826} & \emph{0.826} & \emph{0.826}\\
Fully Independent & 3721 & 3719 & 3717 & 0.761 & 0.761 & 0.761\\
Independent Residuals & \emph{3706} & \emph{3704} & \emph{3703} & 0.759 & 0.759 & 0.759\\
\bottomrule
\end{tabular}
\end{table*}

In terms of MAE, RMSE and CRPS, the Fully Independent model is superior across the board, with considerably worse behavior by the full model and somewhat closer performance from the Independent Residuals model.
This result runs counter to those displayed for wind power production shown in Part 1.  It appears that imposing a correlation structure in the modeling adds spurious dependence.

\begin{table*}[ht]
\centering
\caption{Marginal predictive performance of models for sun power production as measured by mean absolute error (MAE), root mean squared error (RMSE) and mean continuous ranked probability score (CRPS). The results are aggregated over lead times of 1-24h (Day 1), 25-48h (Day 2) and 49-72h (Day 3), and the test period from January 1, 2011 to December 31, 2012. The best results in each category are indicated in italics.}\label{tab:marginal scores}
\vspace{1mm}
\begin{tabular}{lrrrrrrrrr}
\toprule
& \multicolumn{3}{c}{MAE (MW)} & \multicolumn{3}{c}{RMSE (MW)} & \multicolumn{3}{c}{CRPS (MW)} \\
\cmidrule(lr){2-4} \cmidrule(lr){5-7} \cmidrule(lr){8-10} 
& Day 1 & Day 2 & Day 3 & Day 1 & Day 2 & Day 3 & Day 1 & Day 2 & Day 3 \\
\midrule
Full Model & 457 & 452 & 487 & 1047 & 1066 & 1108 & 389 & 396 & 409\\
Fully Independent & \emph{455} & \emph{448} & \emph{487} & \emph{974} & \emph{1007} & \emph{1056} & \emph{344} & \emph{354} & \emph{370}\\
Independent Residuals & 461 & 454 & 498 & 987 & 1021 & 1067 & 346 & 357 & 374\\
\bottomrule
\end{tabular}
\end{table*}

\subsection{Multivariate calibration}
\begin{figure}[!hbpt]
\centering
\includegraphics[width=0.25\textwidth]{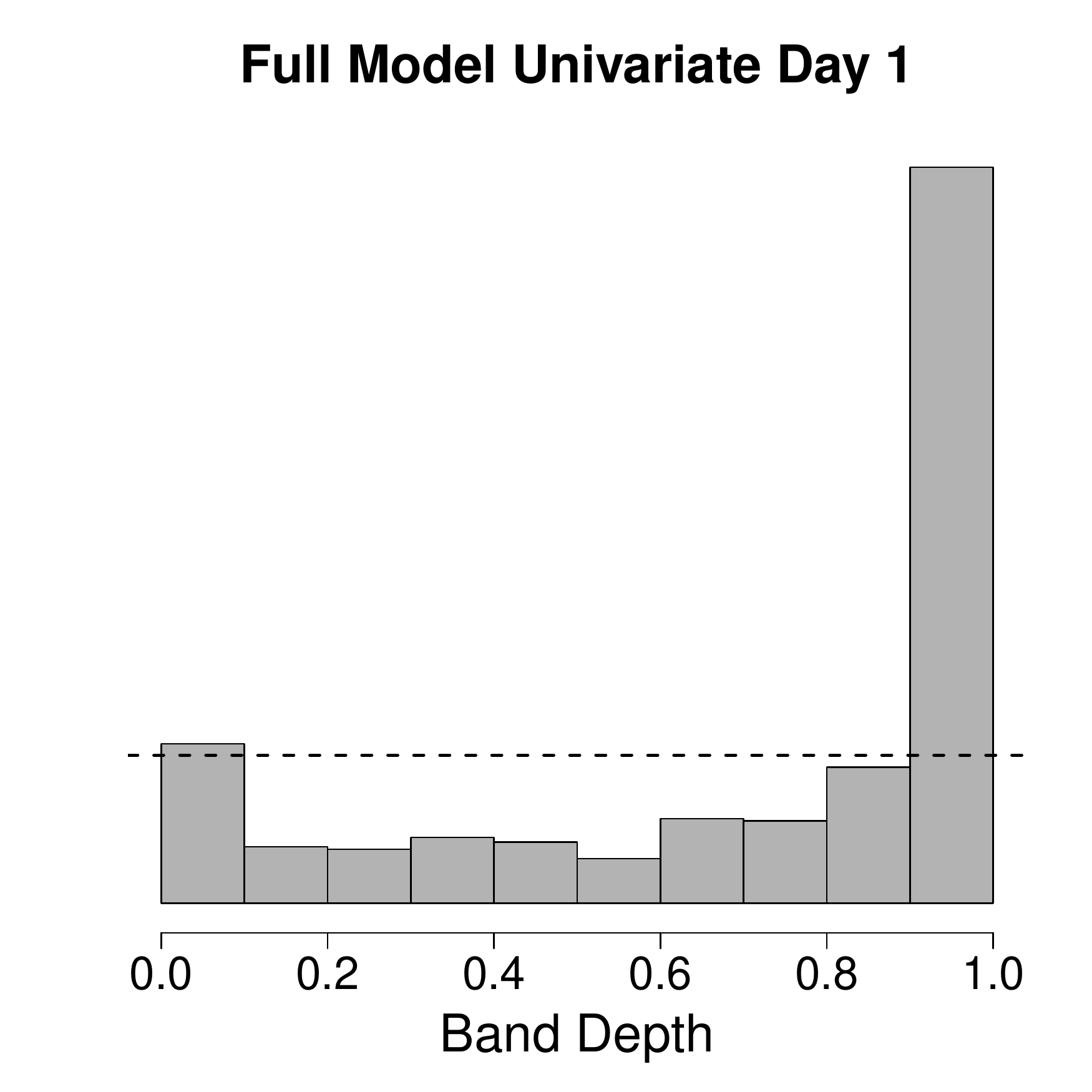}
\includegraphics[width=0.25\textwidth]{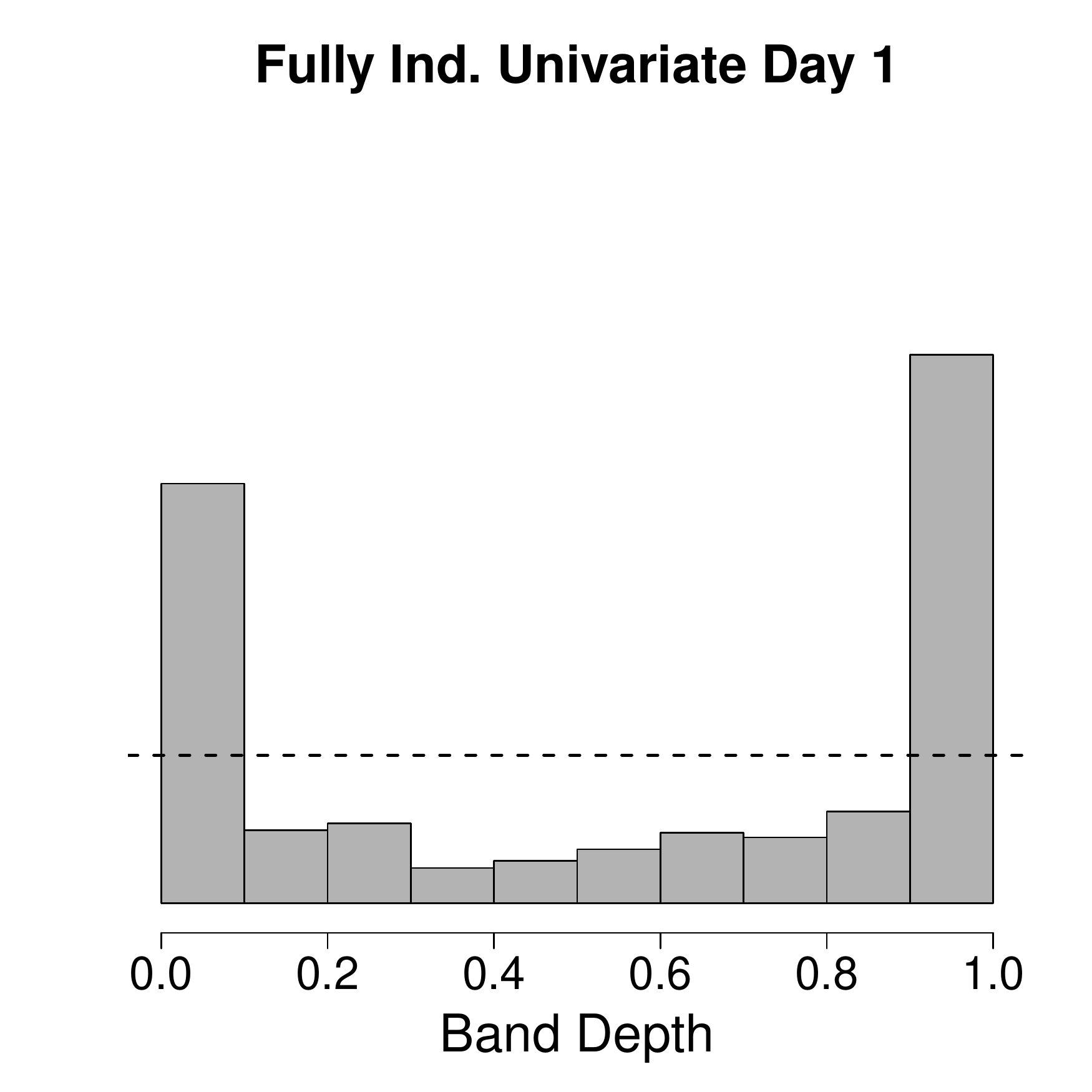}
\includegraphics[width=0.25\textwidth]{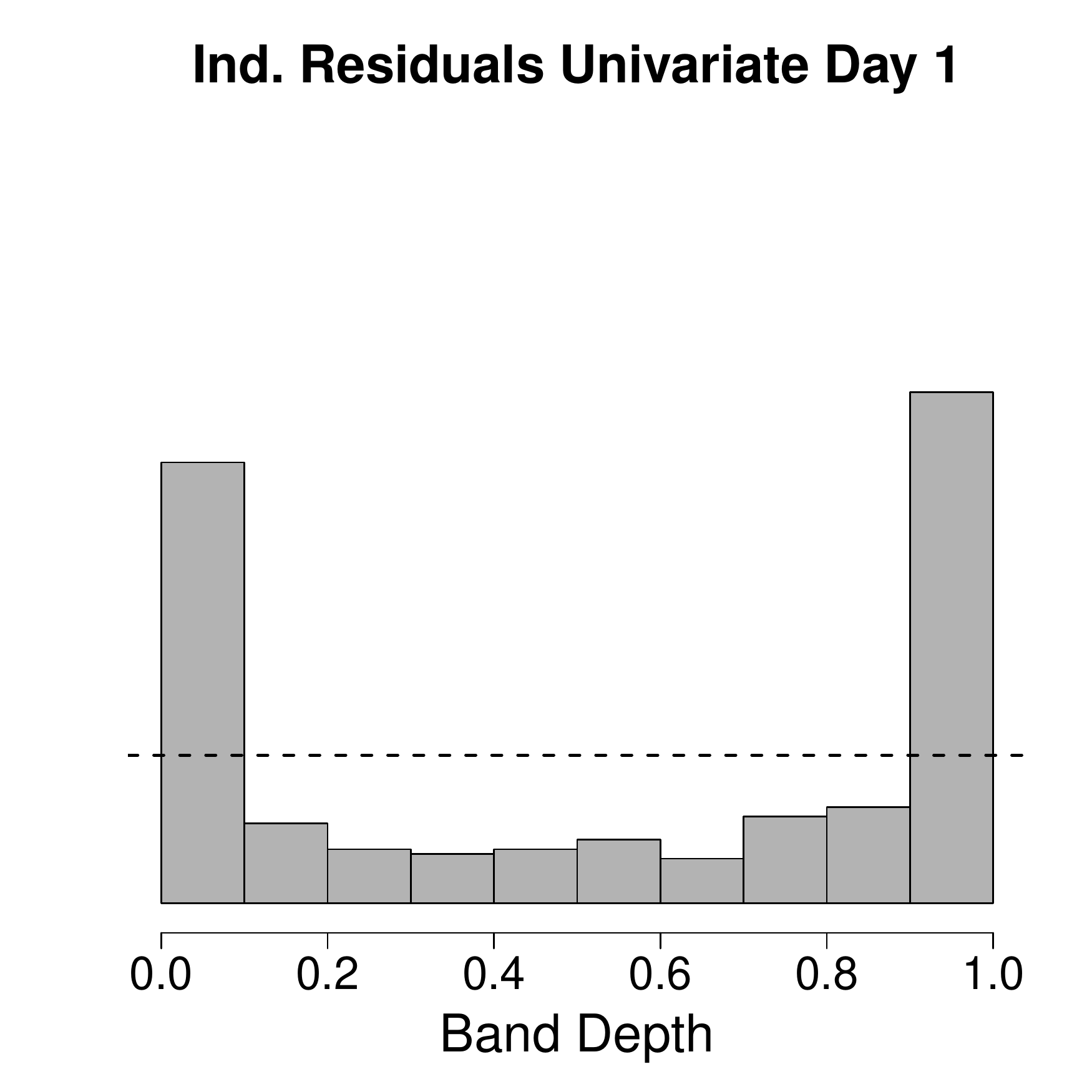}\\
\includegraphics[width=0.25\textwidth]{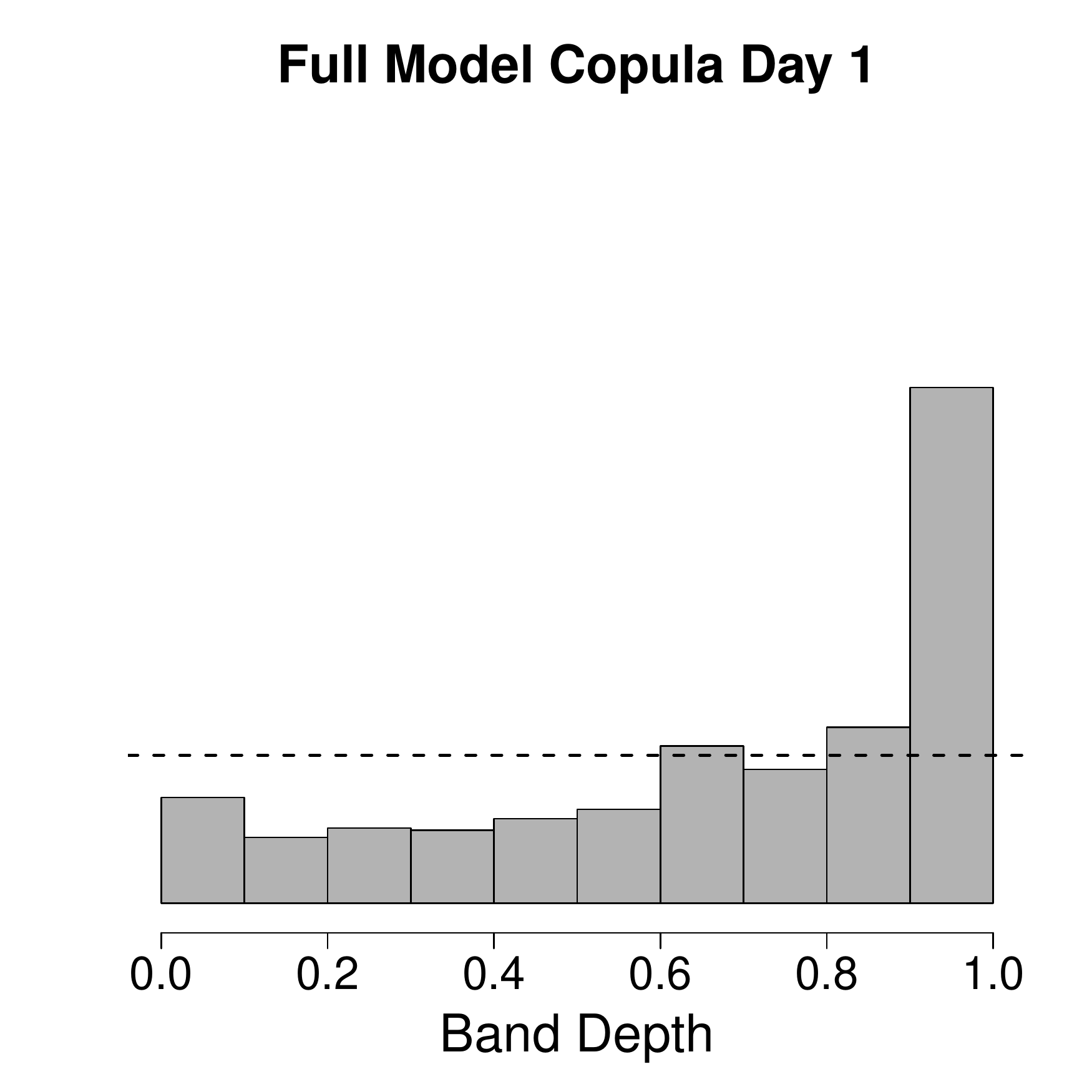}
\includegraphics[width=0.25\textwidth]{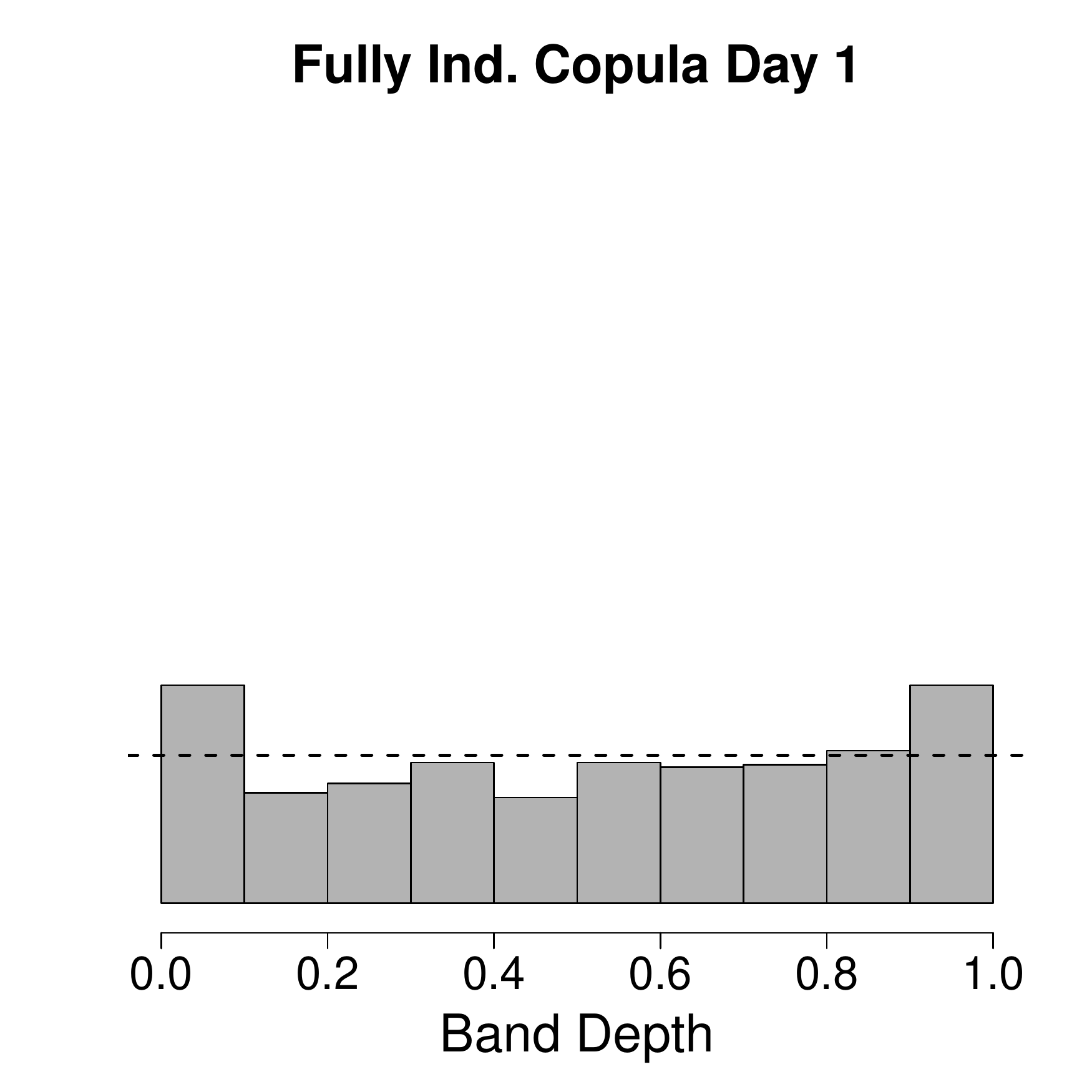}
\includegraphics[width=0.25\textwidth]{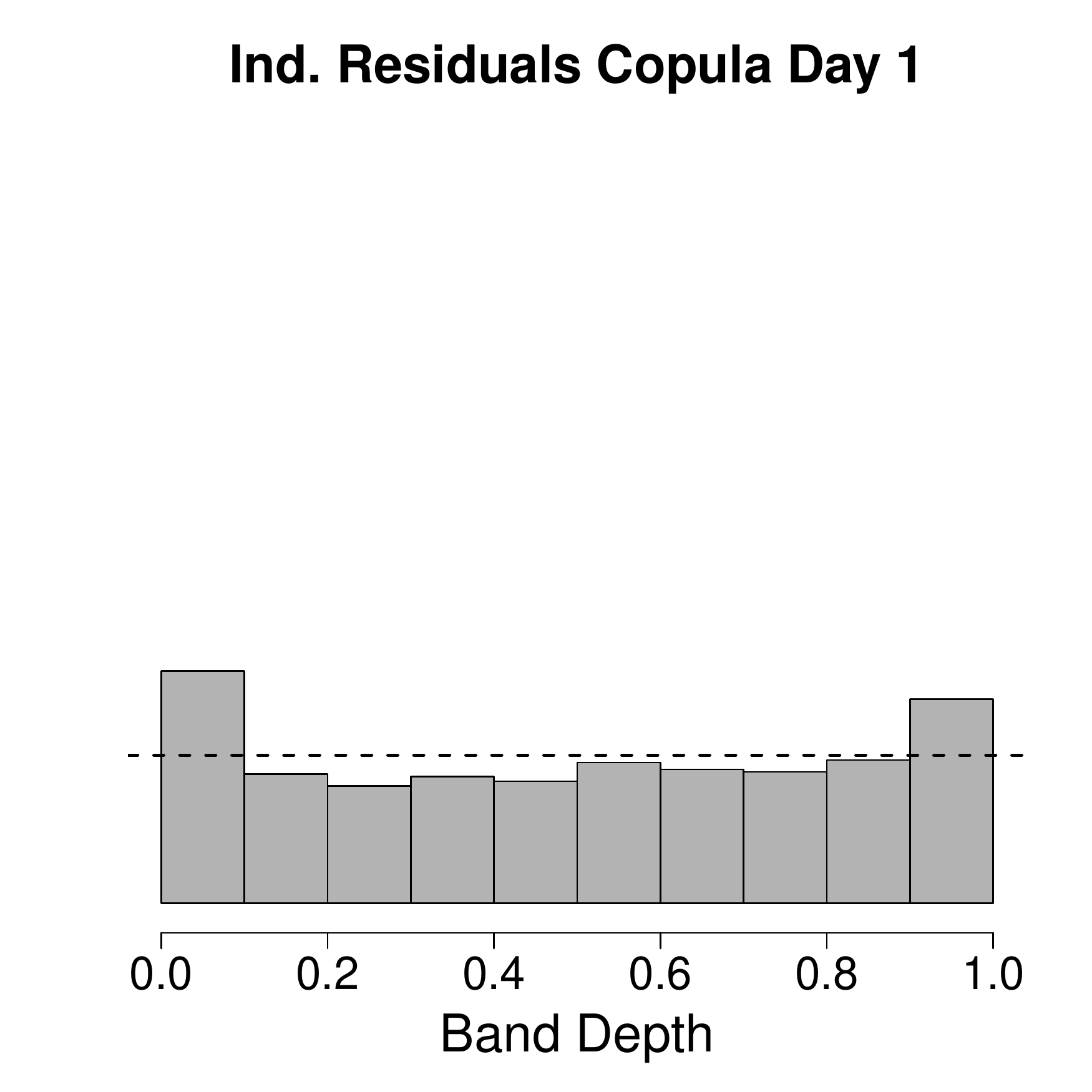}
\caption{Band depth rank histograms for hours 1-24 by model type under either the univariate (top row) or copula (bottom row) approach.}\label{fig:bd_sun}
\end{figure}

Figure~\ref{fig:bd_sun} shows the band depth rank histograms for the joint predictive distribution over hours 1-24.  For each of the three approaches, the independent model is shown on the top row while the bottom row shows the results after copula post-processing.

Figure~\ref{fig:bd_sun} shows two interesting features.  First, the Full Model shows substantial underdispersion and copula post processing does not help this matter.  Secondly, in the case of the Fully Independent and Independent Residuals models the univariate fits show a lack of multivariate calibration which occurs when the predictive distribution is either too focused or too dispersed relative to the true trajectory.  In both cases, copula post-processing nearly eliminates this feature.  This corroborrates the result in Part 1 that marginal modeling with copula post-processing is an effective way of achieving a sharp and calibrated multivariate predictive distribution.

\subsection{Predicting daily maxima and totals}

We now consider the total PV solar production and maximum hourly PV production over the 72 hours.  Since these two quantities are affected by the joint behavior of the underlying forecast, assessments of their distributional performance provides an indication of the quality of the overall joint distributional forecast.
\begin{table*}[htp]
\centering
\caption{Scores for predicting the sum of 72-hours ahead production of PV solar power by method.}\label{tab:mult_sum}
\vspace{1mm}
\begin{tabular}{lrrrrrrrrr}
\toprule
& MAE (MW) & RMSE (MW) & CRPS (MW)\\
\midrule
Full Model Univariate & 24650 & 36975 & 18587\\
Fully Independent Univariate & 23829 & 32514 & 18787\\
Independent Residuals Univariate & 24702 & 33615 & 19536\\
Full Model Copula & 24341 & 37064 & 18962\\
Fully Independent Copula & \emph{23443} & \emph{32404} & \emph{17006}\\
Independent Residuals Copula & 24382 & 33598 & 17531\\
\bottomrule
\end{tabular}
\end{table*}

Table~\ref{tab:mult_sum} shows the scores for each method for the sum of wind power across all 72-hours.  We see that acording to all methodologies the fully independent model with a copula post-processing shows the best performance.  Inn all cases, copula post processing improved the final model score.

\begin{table*}[ht]
\centering
\caption{Scores for predicting the maximum of 72-hours ahead production of PV solar power by method.}\label{tab:mult_max}
\vspace{1mm}
\begin{tabular}{lrrrrrrrrr}
\toprule
& MAE (MW) & RMSE (MW) & CRPS (MW)\\
\midrule
Full Model Univariate & 5210 & 8270 & 3734\\
Fully Independent Univariate & 4312 & 6245 & 3326\\
Independent Residuals Univariate & 4296 & 6279 & 3308\\
Full Model Copula & 3543 & 6495 & 2507\\
Fully Independent Copula & \emph{2195} & \emph{3593} & \emph{1600}\\
Independent Residuals Copula & 2209 & 3699 & 1616\\
\bottomrule
\end{tabular}
\end{table*}

Table~\ref{tab:mult_max} shows similar scores for the maximum.  The conclusions here are broadly in-line with those from Table~\ref{tab:mult_sum}. We see that the fully independent errors model with copula performs best.  In the case of the maximum this improvement is dramatic, with a roughly 50\% reduction in each score.  This speaks the importance of modeling the joint error behavior, but also shows that copula post-processing is sufficient to capture these dependencies.

\begin{figure}[!hbpt]
\centering
\includegraphics[width=0.25\textwidth]{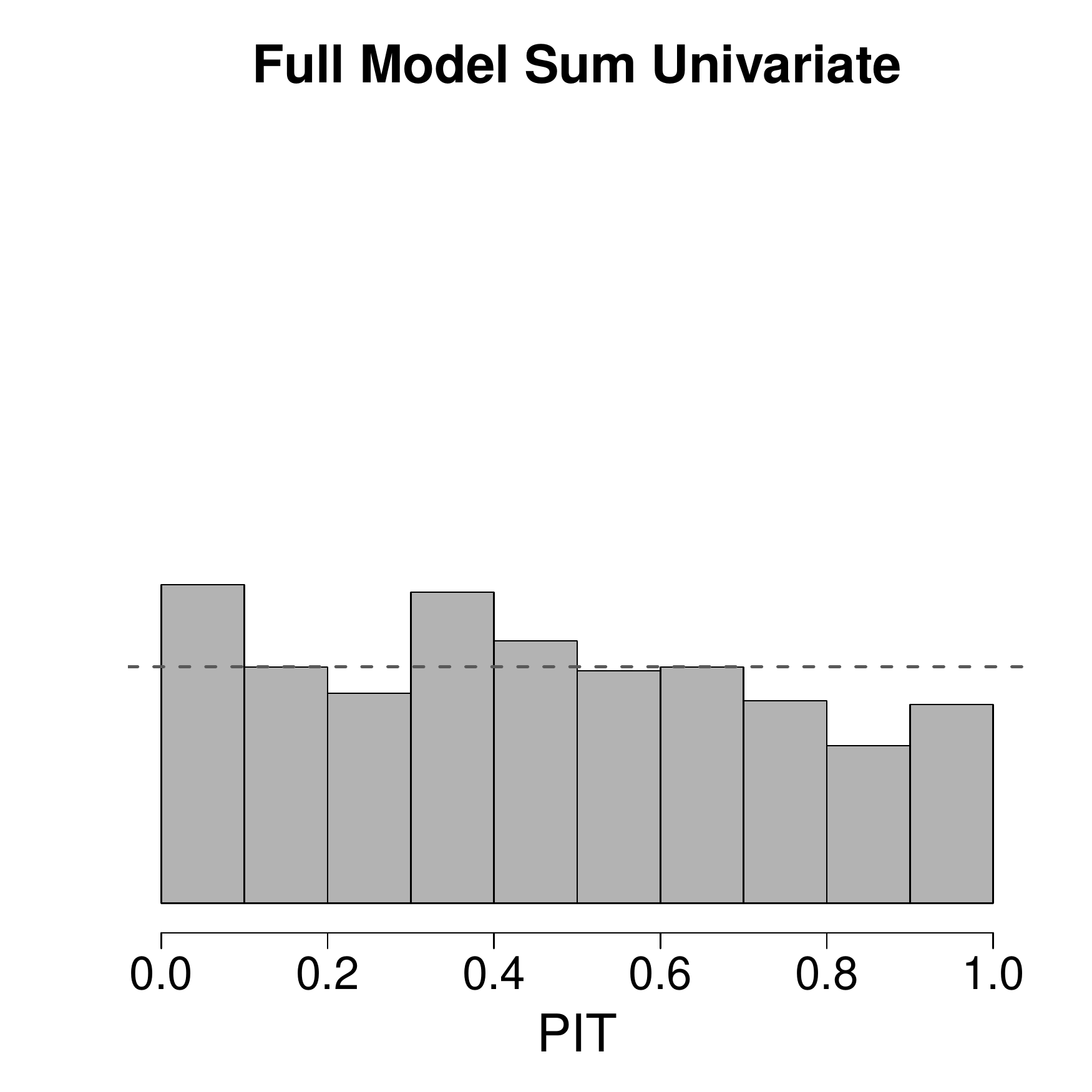}
\includegraphics[width=0.25\textwidth]{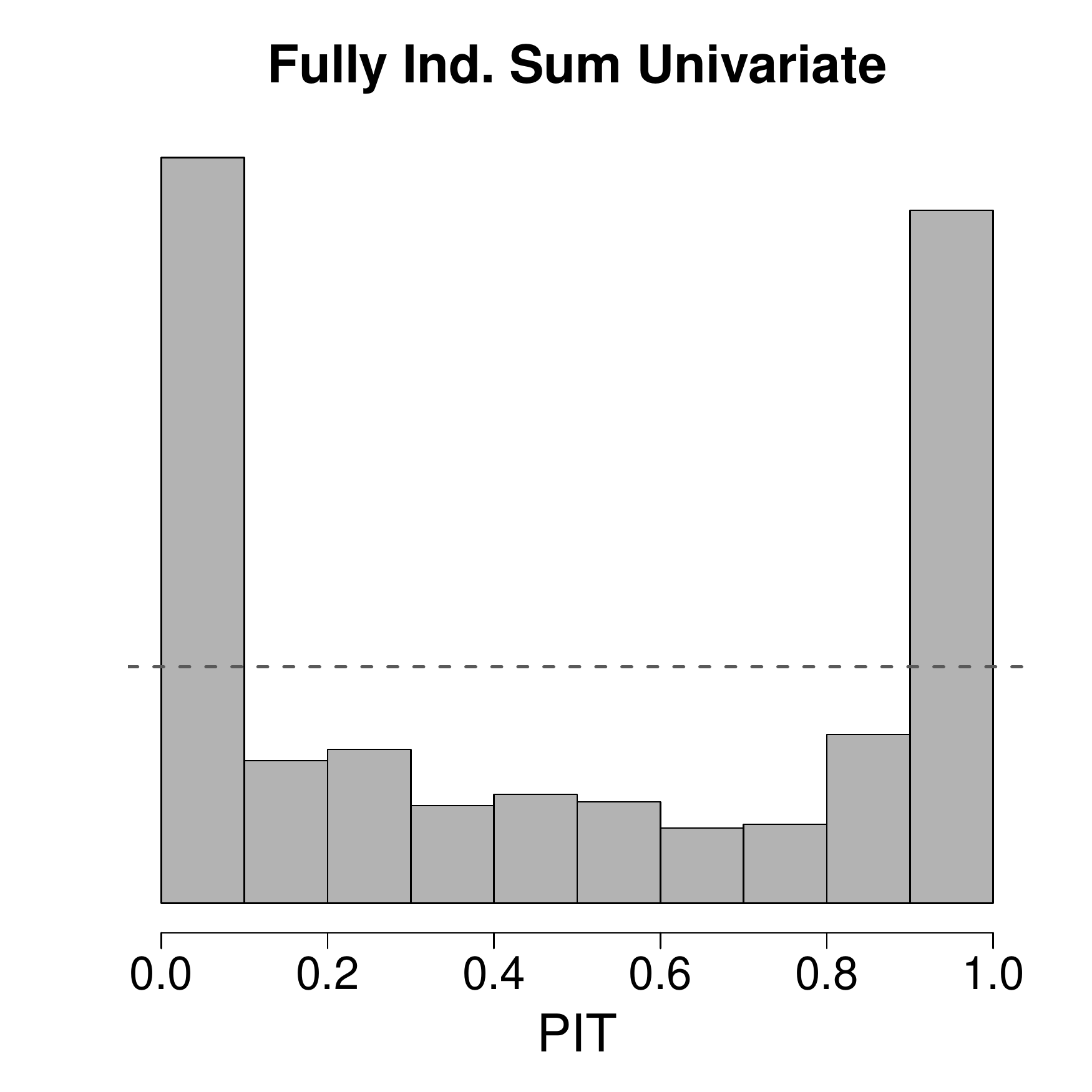}
\includegraphics[width=0.25\textwidth]{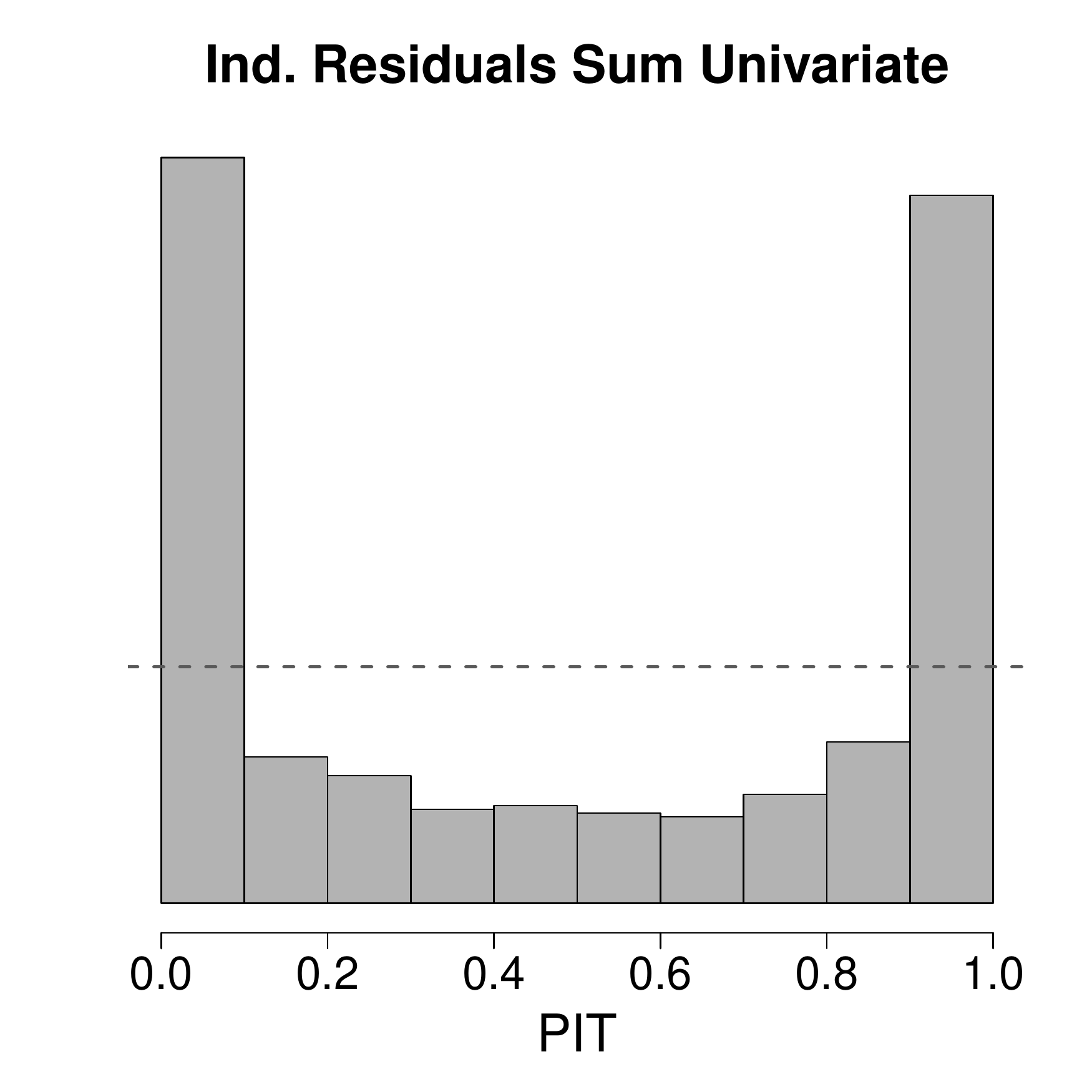}\\
\includegraphics[width=0.25\textwidth]{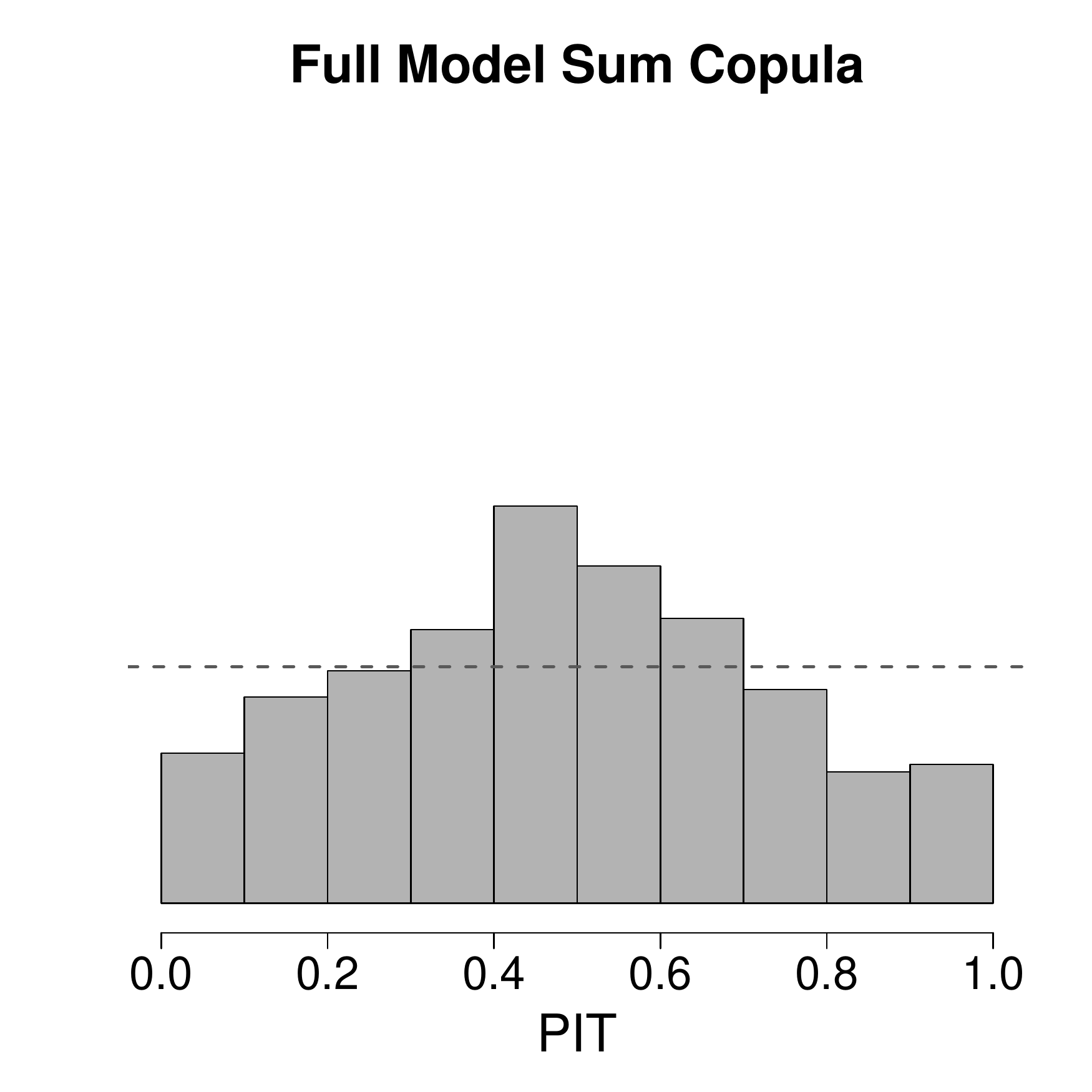}
\includegraphics[width=0.25\textwidth]{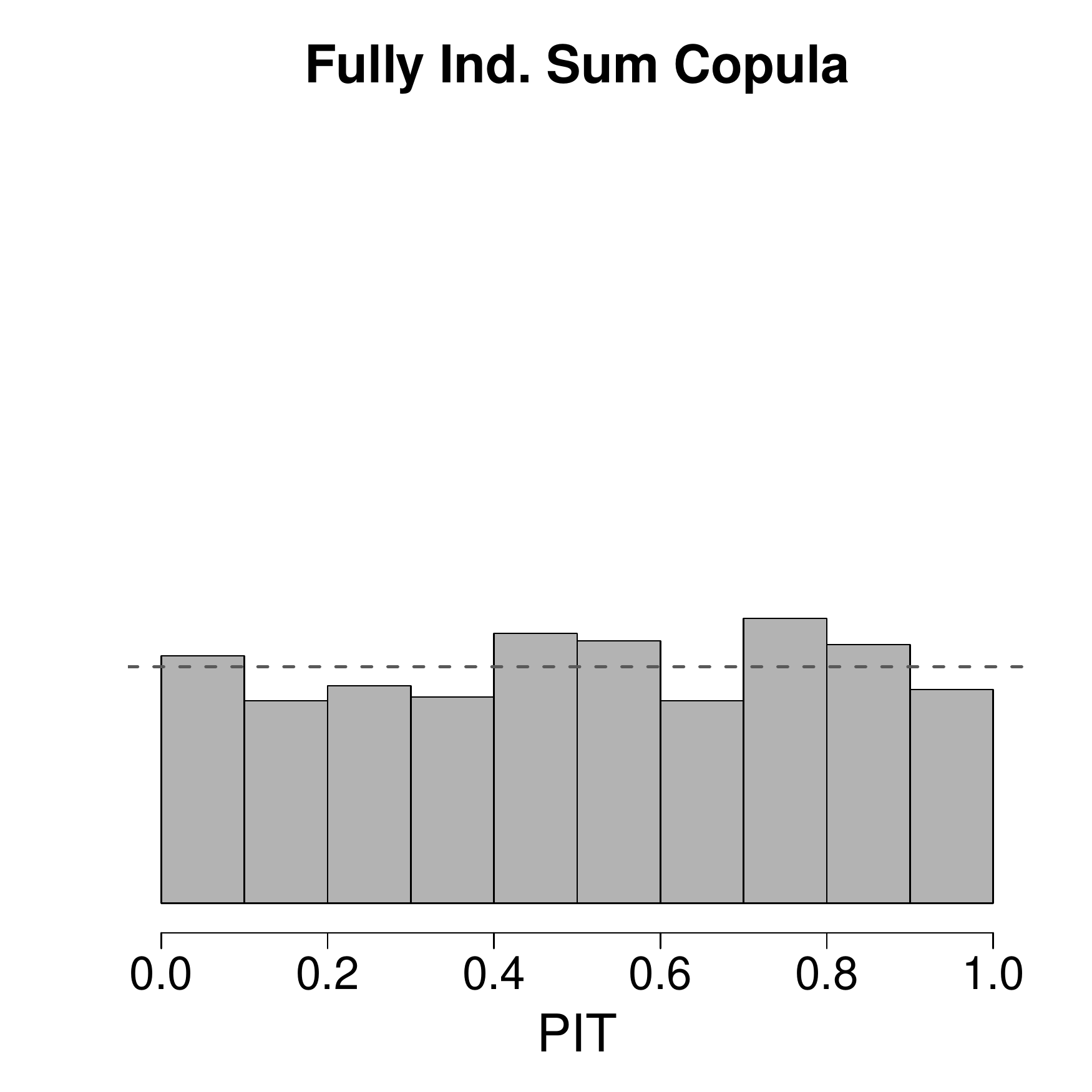}
\includegraphics[width=0.25\textwidth]{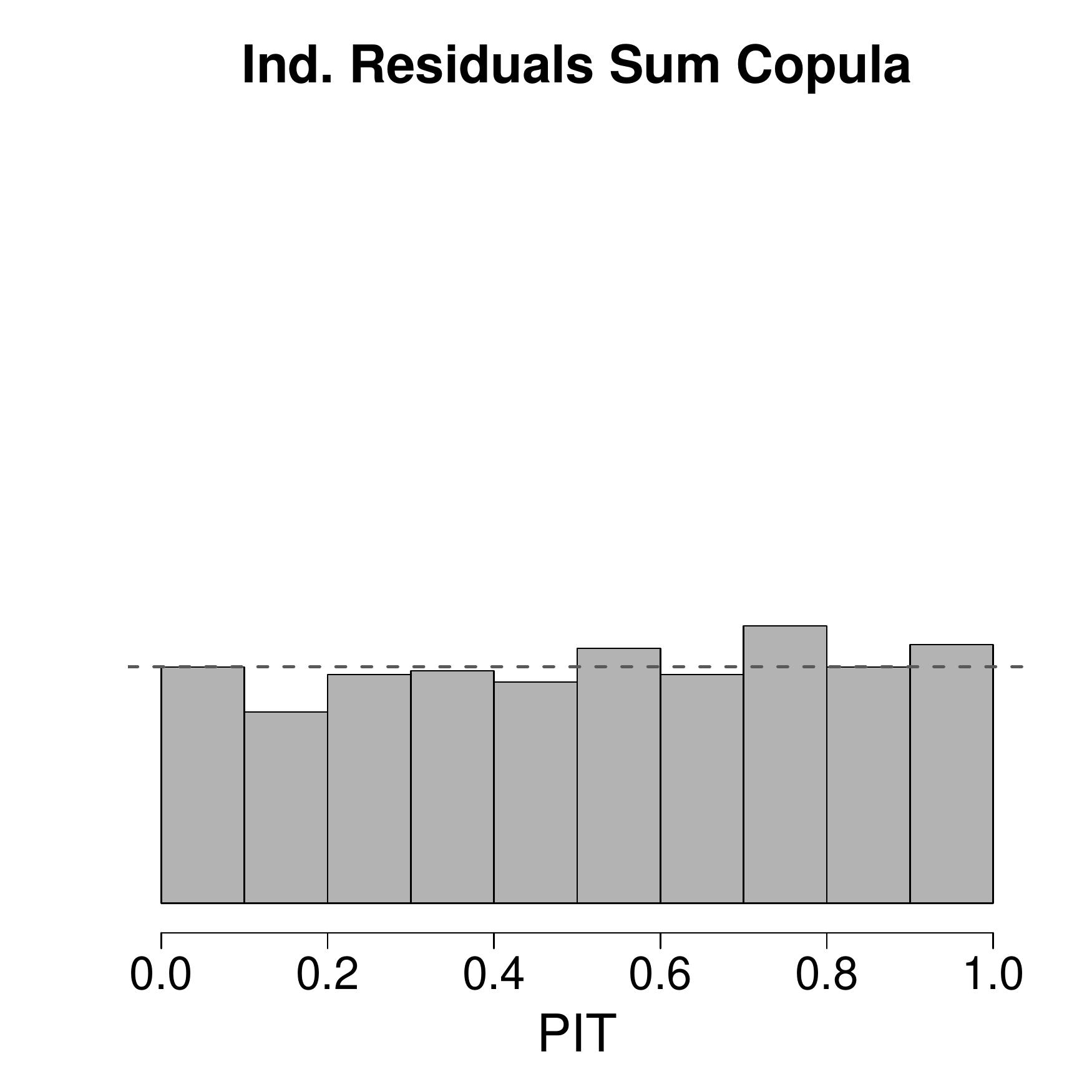}
\caption{PIT for the sum by model type (full, independent errors and fully independent respectively going from left to rigth) and under either the univariate (top row) or copula (bottom row) approach.}\label{fig:hist_sum_sun}
\end{figure}

\begin{figure}[!hbpt]
\centering
\includegraphics[width=0.25\textwidth]{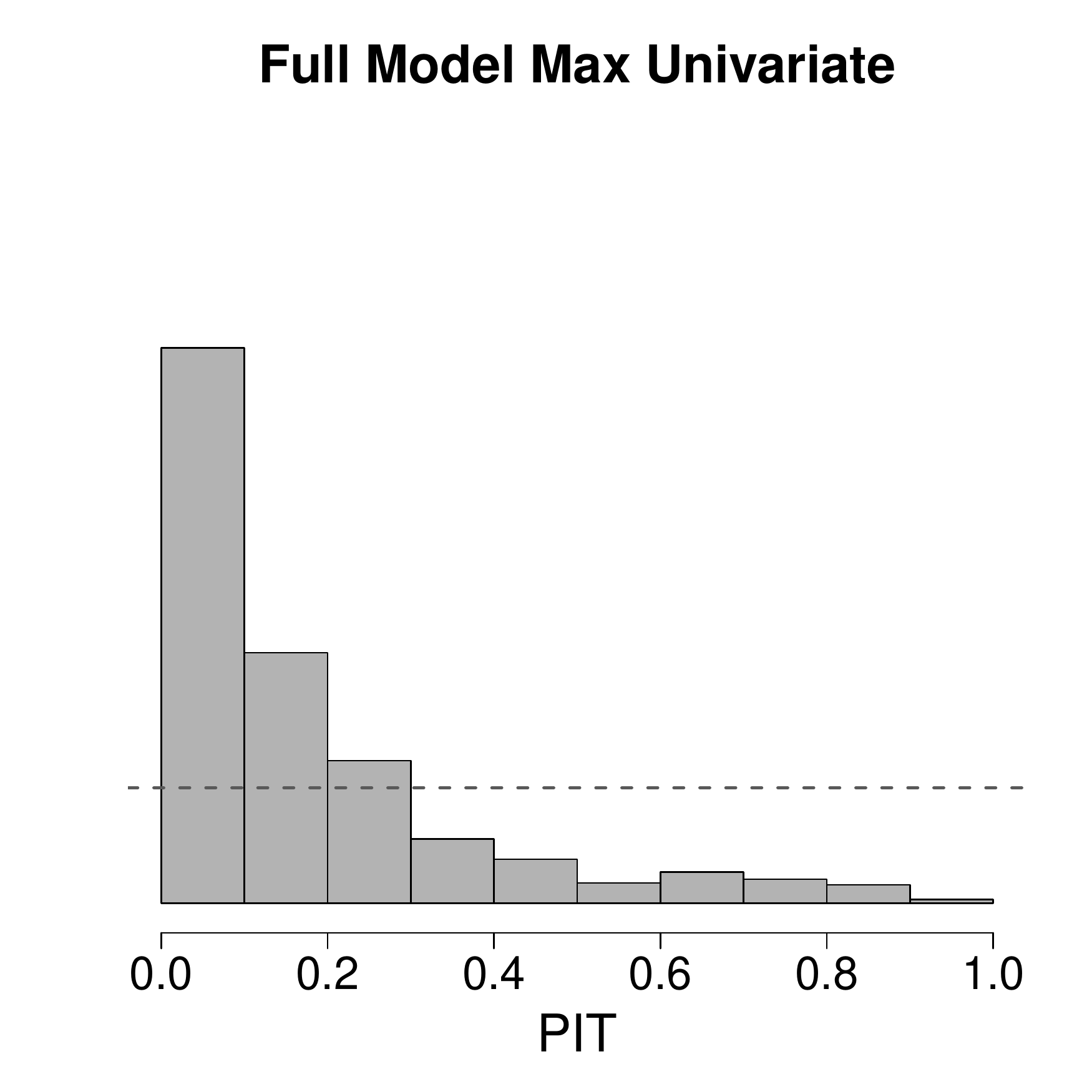}
\includegraphics[width=0.25\textwidth]{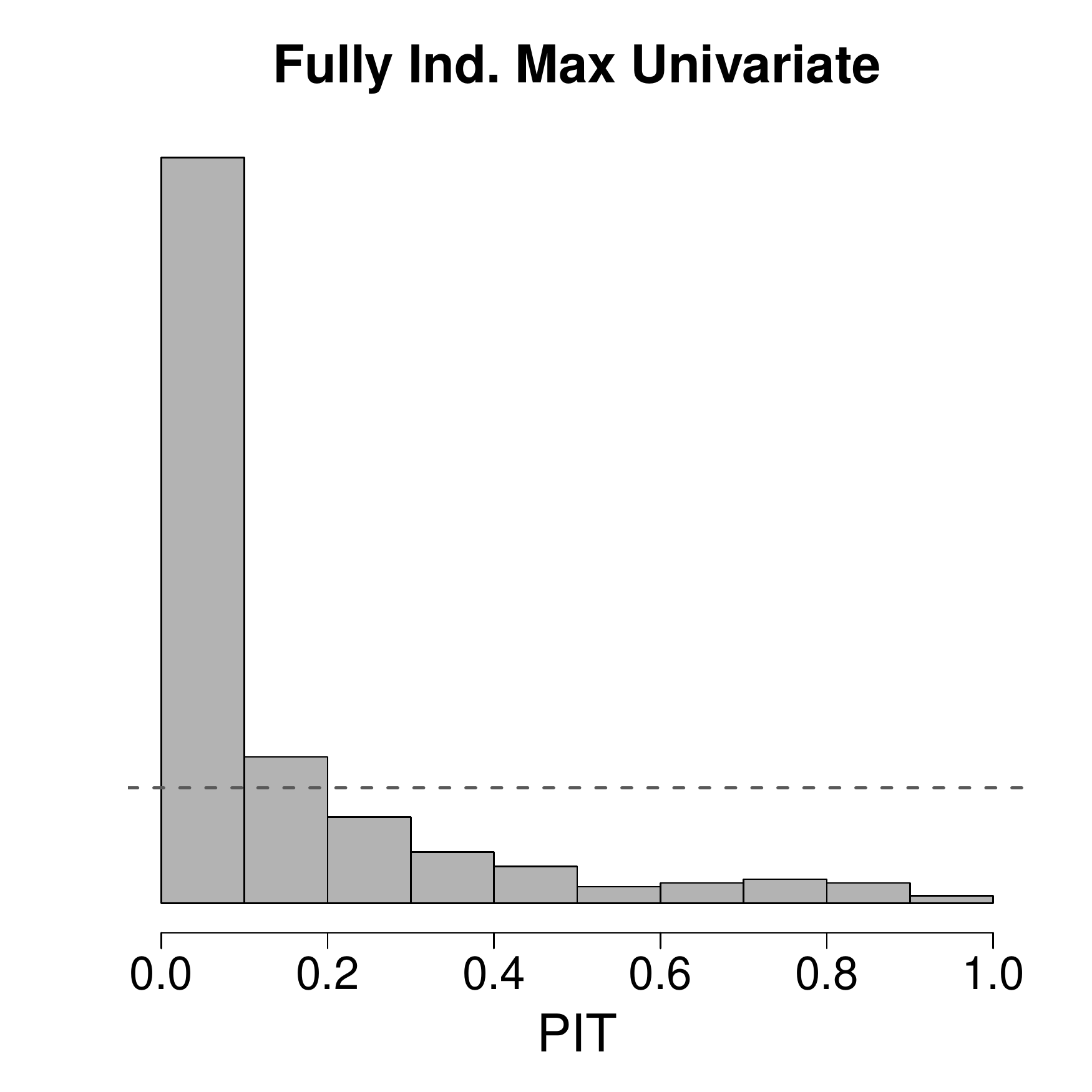}
\includegraphics[width=0.25\textwidth]{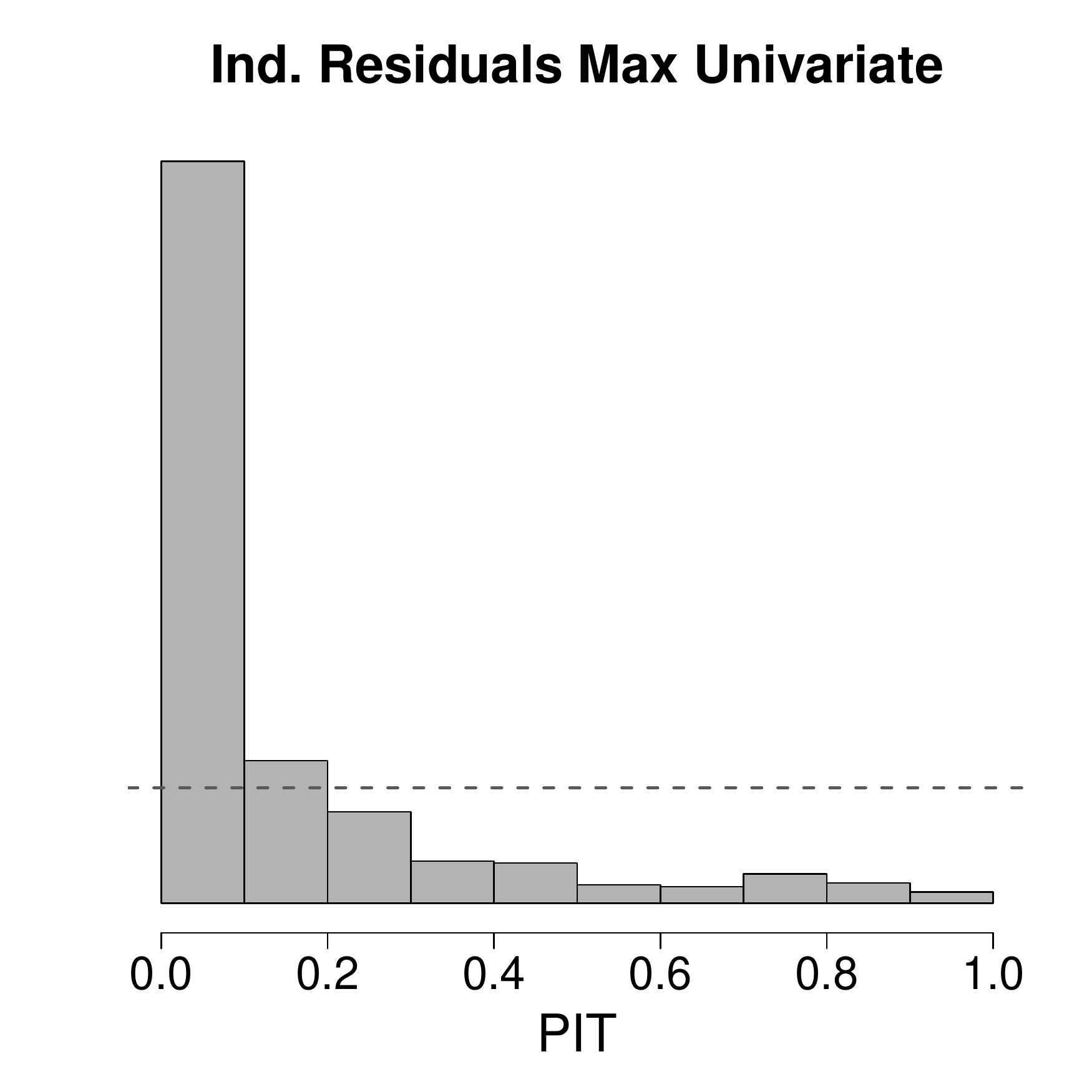}\\
\includegraphics[width=0.25\textwidth]{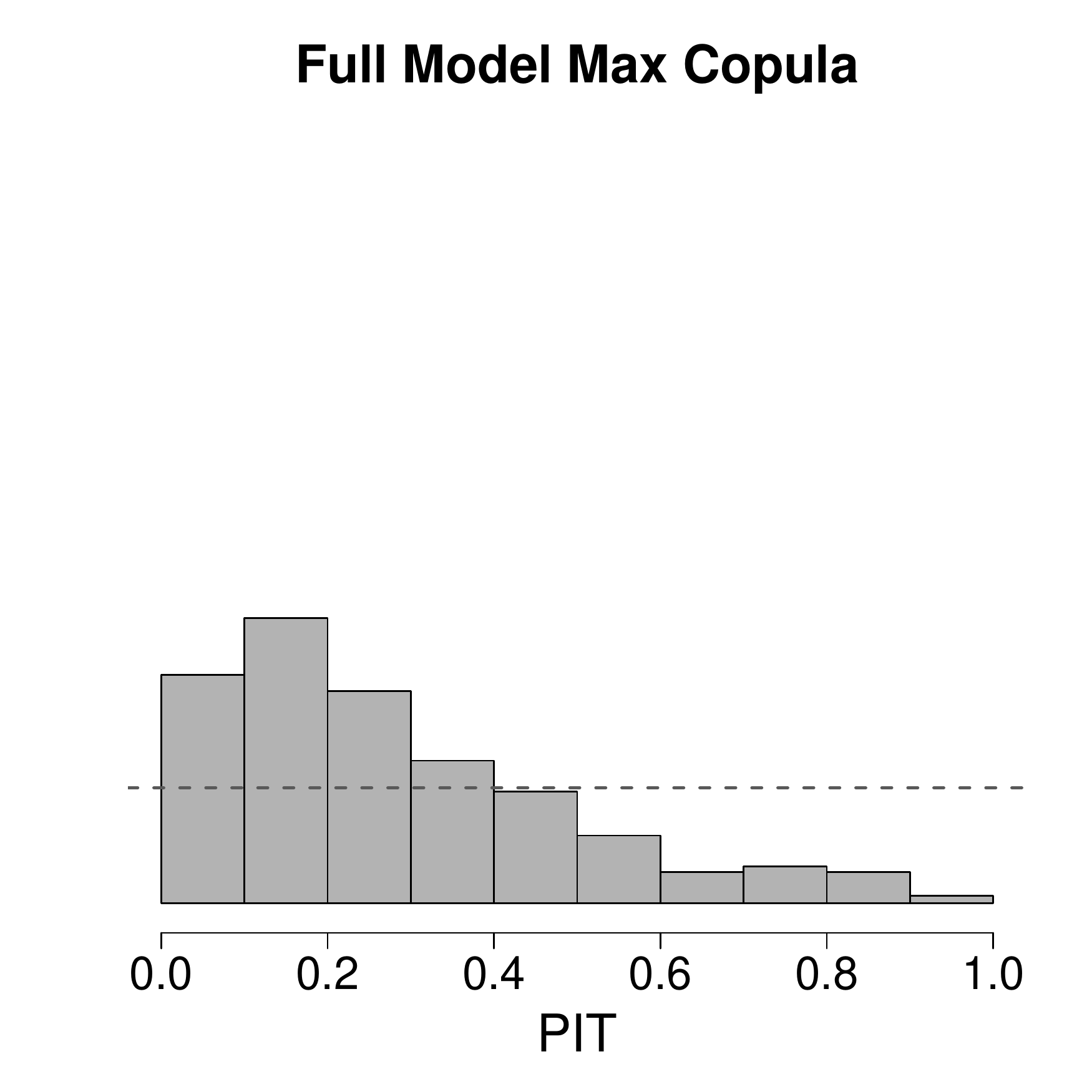}
\includegraphics[width=0.25\textwidth]{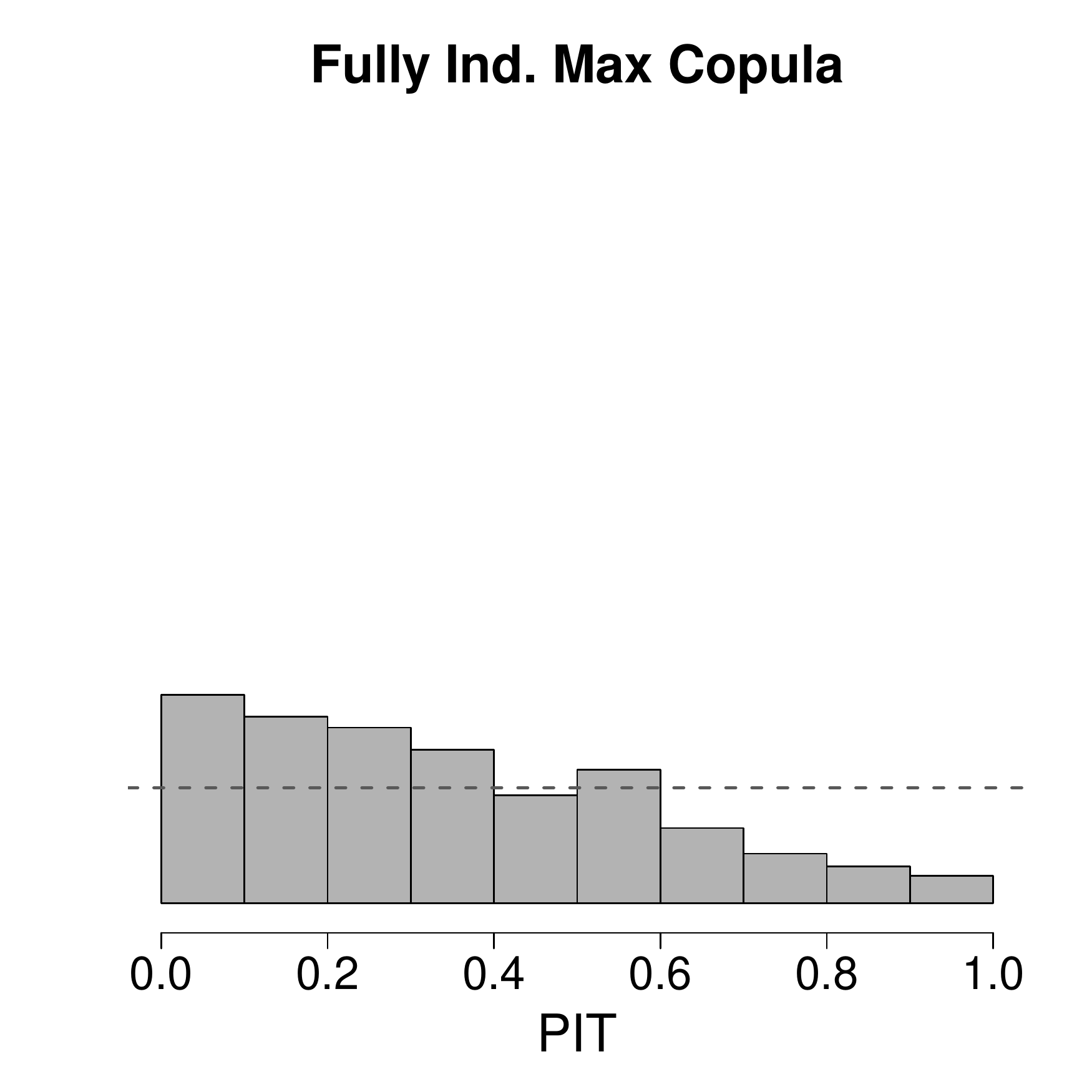}
\includegraphics[width=0.25\textwidth]{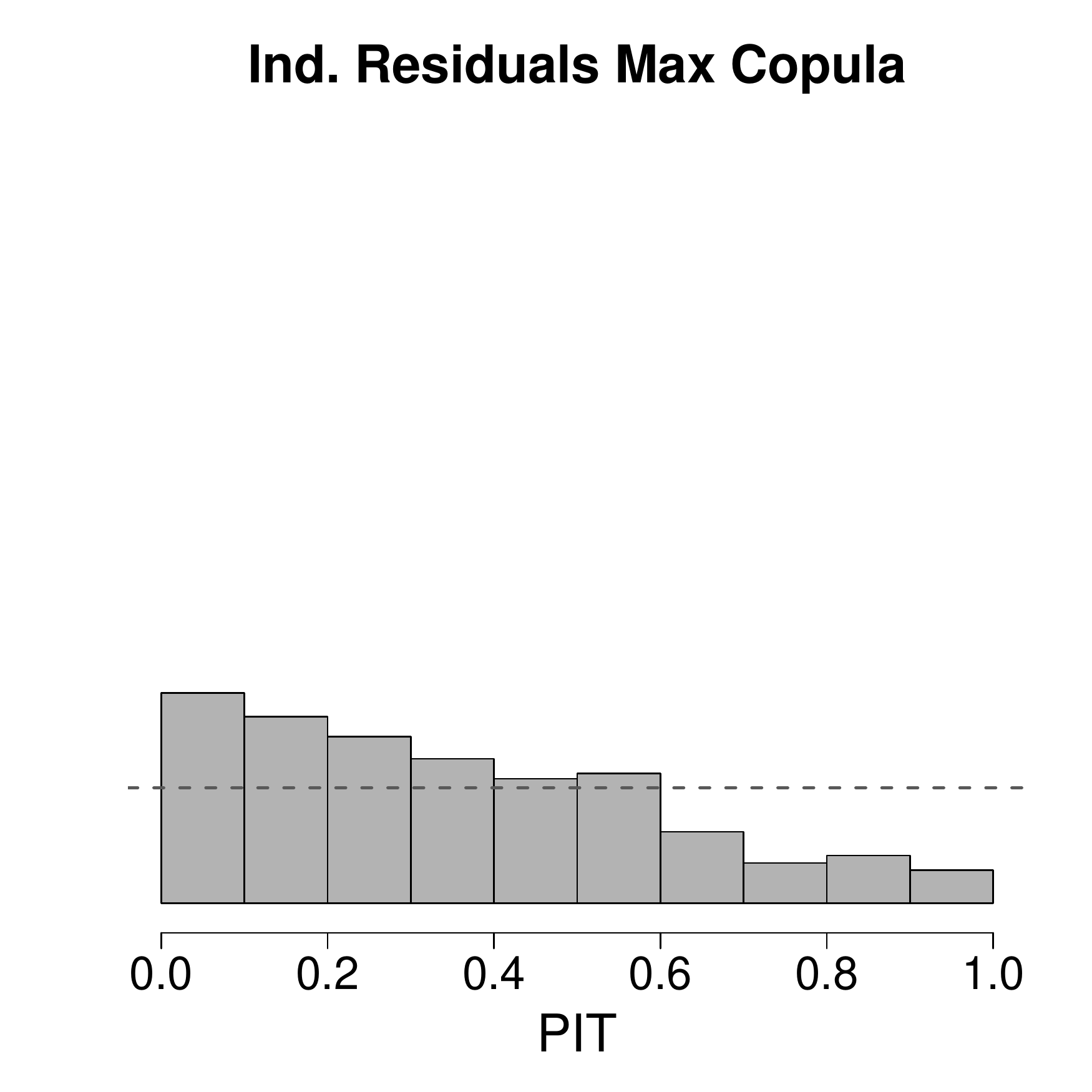}
\caption{PIT for the max by model type (full, independent errors and fully independent respectively going from left to rigth) and under either the univariate (top row) or copula (bottom row) approach.}\label{fig:hist_max_sun}
\end{figure}

Figure~\ref{fig:hist_sum_sun} shows the PIT histograms for the total PV production under the various model combinations.  We see again that while the full model is originally rather calibrated (and indeed becomes less calibrated after a subsequent copula post processing), the best performing models are the fully independent and independent residuals models after a subsequent copula post processing.

Figure~\ref{fig:hist_max_sun} shows the PIT histograms for the maximum PV production over 72 hours. In this case, we see that in all models the predictive distributions are heavily upwards biased. This feature is considerably improved (though not completely mitigated) via copula post processing. 

In general, the results in Tables~\ref{tab:mult_sum}~and~\ref{tab:mult_max} alongside the calibration results in Figures~\ref{fig:hist_sum_sun}~and~\ref{fig:hist_max_sun} show that our approach to joint distributional modeling works well.  Interestingly and in contrast to the results in Part 1, modeling dependence in the regression coefficients did not appear to be beneficial.  However, a post processing step via a Gaussian copula to correlated samples was highly-beneficial and indeed considerably more useful than directly incorporating dependence into a joint model.  The predictive distributions for the maximum PV production are still slightly prone to upward bias, even after copula post-processing, an aspect which constitutes further research.

\section{Discussion\label{sec:discussion}}

We have outlined a model for probabilistic forecasting of hourly PV production in Germany for lead times up to three days ahead.  Our approach used summary measures taken from NWP output as an input feature in a Bayesian regression model.  We entertained two sources of dependence, hierarchical dependence between regression coefficients and residual dependence amongst forecast errors.  It was found that for PV forecasting, hierarchical dependence was not useful and -- in line with the results of Part 1 -- that it is better to impose residual dependence in a second copula post processing step.

These results point to an important aspect related to the training of Bayesian models for use in practical forecasting problems.  In particular, while it is appealing to believe that a single, large joint model can handle all potential dependencies in a system, the potential for model misspecification grows in tandem with the model complexity.  Indeed, as seen here, the Full Model performed far worse than more parsimonious alternatives. Part of the reason, we believe, that the copula post-processing improved model performance was because it uses a second data layer (the actual observed residuals) to incorporate model misspecification into the residual process.  We feel this is an important practical lesson for all forecasters to keep in mind.

The NWP forecast accuracy for GHI is approximately constant for lead times up to 3 days \cite{Remund&2008}, with a significantly better performance under clear skies than for cloudy conditions \cite{Lara-Fanego&2012}. As a result, the predictive performance and the associated forecast uncertainty of the PV power production forecasts is approximately constant for all lead times up to 72 hours. This was not the case for the wind power production forecasts in Part 1, where the forecast uncertainty increased and the predictive performance decreased with longer lead times.   

One final challenge faced by our approach relates to modeling the ``shoulder'' hours of a given day throughout the year.  For example, the PV power production in hour 8 at the start of spring moves rapidly up from 0 (throughout the winter) to a positive number. In these cases, historical training data is not particularly useful, which partially informed the use of a short training period of 20-days.  A more direct model which incorporated this seasonality should be able to capture these dynamics and thereby enable a wider training window to be entertained.

Parts 1 and 2 of this considered renewable energy production from wind and solar power seperately.  Clearly, forecasts of their joint production is of interest to actors in power markets. The methodologies we have outlined, especially the copula post-processing steps should be capable of accommodating these larger forecasting objectives.  Production across multiple countries can be entertained in a similar manner.  In both cases, the graphical conditional independence structure that was imposed in the copula model should prove a useful means of avoiding over-specification.

\section*{Acknowledgment}
This work was performed within Big Insight -- Centre for Research-based Innovation with support from The Research Council of Norway through grant nr. 237718. We thank Stefan Erath from Norsk Hydro for sharing his expertise and data.


\begin{thebibliography}{10}

\bibitem{agoua2018probabilistic}
X~G Agoua, R Girard, and G Kariniotakis.
\newblock Probabilistic model for spatio-temporal photovoltaic power
  forecasting.
\newblock {\em {IEEE Transactions on Sustainable Energy}}, 2018.

\bibitem{alessandrini2015analog}
S~Alessandrini, L~Delle~Monache, S~Sperati, and G~Cervone.
\newblock An analog ensemble for short-term probabilistic solar power forecast.
\newblock {\em {Applied Energy}}, 157:95--110, 2015.

\bibitem{AndrewsPearce2012}
R~W Andrews and J~M Pearce.
\newblock Prediction of energy effects on photovoltaic systems due to snowfall
  events.
\newblock In {\em Photovoltaic Specialists Conference (PVSC), 2012 38th IEEE},
  pages 003386--003391. IEEE, 2012.

\bibitem{bacher2009}
P~Bacher, H~Madsen, and H~A Nielsen.
\newblock Online short-term solar power forecasting.
\newblock {\em {Solar Energy}}, 83(10):1772--1783, 2009.

\bibitem{Bracale&2013}
A~Bracale, P~Caramia, G~Carpinelli, and A~R {Di Fazio}.
\newblock A {B}ayesian method for short-term probabilistic forecasting of
  photovolataic generation in smart grid operation and control.
\newblock {\em Energies}, 6:733--747, 2013.

\bibitem{Bremnes2004}
J~B Bremnes.
\newblock Probabilistic wind power forecasts using local quantile regression.
\newblock {\em Wind Energy}, 7:47--54, 2004.

\bibitem{Dawid1984}
A~P Dawid.
\newblock Statistical theory: The prequential approach (with discussion and
  rejoinder).
\newblock {\em Journal of the Royal Statistical Society Ser.~A}, 147:278--292,
  1984.

\bibitem{Dowell&2016}
J~Dowell and P~Pinson.
\newblock Very-short-term probabilistic wind power forecasts by sparse vector
  autoregression.
\newblock {\em IEEE Transactions on Smart Grid}, 7(2):763--770, 2016.

\bibitem{Drews&2007}
A~Drews, A~C {de Keizer}, H~G Beyer, E~Lorenz, J~Betcke, W~G J H~M {van Sark},
  W~Heydenreich, E~Wiemken, S~Stettler, P~Toggweiler, S~Bofigner, M~Schneider,
  G~Heilscher, and D~Heinemann.
\newblock Monitoring and remote failure detection of grid-connected {PV}
  systems based on satellite observations.
\newblock {\em Solar Energy}, 81:548--564, 2007.

\bibitem{Girodo2006}
M~Girodo.
\newblock {\em Solarstrahlungsvorhersage auf der Basis numerischer
  Wettermodelle}.
\newblock PhD thesis, Dissertation, Universit{\"a}t Oldenburg, 2006.

\bibitem{Gneiting2011}
T~Gneiting.
\newblock Making and evaluating point forecasts.
\newblock {\em Journal of the American Statistical Association}, 106:746--762,
  2011.

\bibitem{GneitingRaftery2007}
T~Gneiting and A~E Raftery.
\newblock Strictly proper scoring rules, prediction, and estimation.
\newblock {\em Journal of the American Statistical Association}, 102:359--378,
  2007.

\bibitem{golestaneh2016very}
F~Golestaneh, P~Pinson, and H~B~Gooi.
\newblock Very short-term nonparametric probabilistic forecasting of renewable
  energy generation—with application to solar energy.
\newblock {\em {IEEE Transactions on Power Systems}}, 31(5):3850--3863, 2016.

\bibitem{Hong&2016}
T~Hong, P~Pinson, S~Fan, H~Zareipour, A~Troccoli, and
  R~J~Hyndman.
\newblock Probabilistic energy forecasting: Global energy forecasting
  competition 2014 and beyond.
\newblock {\em International Journal of Forecasting}, 32(3):896--913, 2016.

\bibitem{Huang&2016}
J~Huang and M~Perry.
\newblock A semi-empirical approach using gradient boosting and k-nearest
  neighbors regression for {GEFCom2014} probabilistic solar power forecasting.
\newblock {\em International Journal of Forecasting}, 32(3):1081--1086, 2016.

\bibitem{JeonTaylor2012}
J~Jeon and J~W Taylor.
\newblock Using conditional kernel density estimation for wind power density
  forecasting.
\newblock {\em Journal of the American Statistical Association},
  107(497):66--79, 2012.

\bibitem{Lara-Fanego&2012}
V~Lara-Fanego, J~A Ruiz-Arias, D~Pozo-V\'azquez, F~J Santos-Alamillos, and
  J~Tovar-Pescador.
\newblock Evaluation of the {WRF} model solar irradiance forecasts in
  {A}ndalusia (southern {S}pain).
\newblock {\em Solar Energy}, 86:2200--2217, 2012.

\bibitem{Lenkoski2013}
A~Lenkoski.
\newblock A direct sampler for {G}-{Wishart} variates.
\newblock {\em Stat}, 9:119--128, 2013.

\bibitem{LeutbecherPalmer2008}
M~Leutbecher and T~N Palmer.
\newblock Ensemble forecasting.
\newblock {\em Journal of Computational Physics}, 227:3515--3539, 2008.

\bibitem{Lorenz&2009}
E~Lorenz, J~Hurka, D~Heinemann, and H~G Beyer.
\newblock Irradiance forecasting for the power prediction of grid-connected
  photovoltaic systems.
\newblock {\em IEEE Journal of Selected Topics in Applied Earth Observations
  and Remote Sensing}, 2(1):2--10, 2009.

\bibitem{Lorenz&2008}
E~Lorenz, J~Hurka, G~Karampela, D~Heinemann, H~G Beyer, and M~Schneider.
\newblock Qualified forecast of ensemble power production by spatially
  dispersed grid-connected pv systems.
\newblock In {\em Proc. of the 23rd European PV Conference}, 2008.

\bibitem{lorenz2011}
E~Lorenz, T~Scheidsteger, J~Hurka, D~Heinemann, and
  C~Kurz.
\newblock Regional {PV} power prediction for improved grid integration.
\newblock {\em Progress in Photovoltaics: Research and Applications},
  19(7):757--771, 2011.

\bibitem{MathiesenKleissl2011}
P~Mathiesen and J~Kleissl.
\newblock Evaluation of numerical weather prediction for intra-day solar
  forecasting in the continental {U}nited {S}tates.
\newblock {\em Solar Energy}, 85(5):967--977, 2011.

\bibitem{Mohammed&2015}
A~A Mohammed, W~Yaqub, and Z~Aung.
\newblock Probabilistic forecasting of solar power: An ensemble learning
  approach.
\newblock In {\em Intelligent Decision Technologies}, pages 449--458. Springer,
  2015.

\bibitem{Molteni&1996}
A~Molteni, R~Buizza, T~N Palmer, and T~Petroliagis.
\newblock The new {ECMWF} ensemble prediction system: Methodology and
  validation.
\newblock {\em Quarterly journal of the royal meteorological society},
  122:73--119, 1996.

\bibitem{nagy&2016}
G~I Nagy, G~ Barta, S~Kazi, G~Borb{\'e}ly, and
  G~Simon.
\newblock {GEFCom2014}: Probabilistic solar and wind power forecasting using a
  generalized additive tree ensemble approach.
\newblock {\em International Journal of Forecasting}, 32(3):1087--1093, 2016.

\bibitem{PellandGalanisKallos2013}
S~Pelland, G~Galanis, and G~Kallos.
\newblock Solar and photovoltaic forecasting through post-processing of the
  {G}lobal {E}nvironmental {M}ultiscale numerical weather prediction model.
\newblock {\em Progress in Photovoltaics: Research and Applications},
  21:284--296, 2013.

\bibitem{Pinson2012}
P~Pinson.
\newblock Very-short-term probabilistic forecasing of wind power with
  generalized logit-normal distributions.
\newblock {\em Journal of the Royal Statistical Society Series C},
  61(4):555--576, 2012.

\bibitem{Pinson2013}
P~Pinson.
\newblock Wind energy: {F}orecasting challenges for its operational management.
\newblock {\em Statistical Science}, 28(4):564--585, 2013.

\bibitem{PinsonKariniotakis2010}
P~Pinson and G~N Kariniotakis.
\newblock Conditional prediction intervals of wind power generation.
\newblock {\em IEEE Transactions on Power Systems}, 25:1845--1856, 2010.

\bibitem{Remund&2008}
J~Remund, R~Perez, and E~Lorenz.
\newblock Comparison of solar radiation forecasts for the {USA}.
\newblock In {\em Proc. of the 23rd European PV Conference}, 2008.

\bibitem{Roverato2002}
A~Roverate.
\newblock Hyper inverse {W}ishart distribution for non-decomposable graphs and
  its applications to {B}ayesian inference for {G}aussian graphical models.
\newblock {\em Scandinavian Journal of Statistics}, 29:391--411, 2002.

\bibitem{sperati2016application}
S~Sperati, S~Alessandrini, and L~D~Monache.
\newblock An application of the ecmwf ensemble prediction system for short-term
  solar power forecasting.
\newblock {\em {Solar Energy}}, 133:437--450, 2016.

\bibitem{Thorarinsdottir&2016}
T~L Thorarinsdottir, M~Scheuerer, and C~Heinz.
\newblock Assessing the calibration of high-dimensional ensemble forecasts
  using rank histograms.
\newblock {\em Journal of Computational and Graphical Statistics},
  25(1):105--122, 2013.

\end{thebibliography}
\end{document}